\documentclass[aps,twocolumn,showpacs,preprintnumbers,floatfix,amssymb]{revtex4} 

\usepackage{epsfig}
\usepackage{graphicx}
\begin{document}

\preprint{
%\vbox{
%\hbox{hep-lat/0211017}
\hbox{ADP-02-97/T535}
%}
}

\title{Quark Contributions to Baryon Magnetic Moments\\ in Full, Quenched
and Partially Quenched QCD}

\author{Derek B. Leinweber}

\email{dleinweb@physics.adelaide.edu.au}
\homepage{http://www.physics.adelaide.edu.au/theory/staff/leinweber/}

\affiliation{Department of Physics and Mathematical Physics and
         Special Research Centre for the Subatomic Structure of Matter,
         University of Adelaide 5005, Australia}

\begin{abstract}
The chiral nonanalytic behaviour of quark-flavor contributions to the
magnetic moments of octet baryons are determined in full, quenched and
partially-quenched QCD, using an intuitive and efficient diagrammatic
formulation of quenched and partially-quenched chiral perturbation
theory.  The technique provides a separation of quark-sector
magnetic-moment contributions into direct sea-quark loop,
valence-quark, indirect sea-quark loop and quenched valence
contributions, the latter being the conventional view of the quenched
approximation.  Both meson and baryon mass violations of
$SU(3)$-flavor symmetry are accounted for.  Following a comprehensive
examination of the individual quark-sector contributions to octet
baryon magnetic moments, numerous opportunities to observe and test
the underlying structure of baryons and the nature of chiral
nonanalytic behavior in QCD and its quenched variants are discussed.
In particular, the valence $u$-quark contribution to the proton
magnetic moment provides the optimal opportunity to directly view
nonanalytic behavior associated with the meson cloud of full QCD and
the quenched meson cloud of quenched QCD.  The $u$ quark in $\Sigma^+$
provides the best opportunity to display the artifacts of the quenched
approximation.
\end{abstract}


\maketitle

\begin{section}{Introduction}
\label{sec:intro}

One of the grand promises of lattice-regularized QCD is to provide
{\it ab initio} predictions of hadronic observables with statistical
and systematic uncertainties constrained to within 1\%.  To realize
this goal, extrapolations are required.  The finite lattice spacing
must be extrapolated to the continuum limit.  The impact of the finite
volume of the periodic lattice must be understood and extrapolated to
infinity.  For computational reasons, the masses of the light $u$ and
$d$ quarks must be extrapolated from rather large values to the point
at which the physical hadron masses are reproduced.

The realization of chiral symmetry in the dynamically-broken Goldstone
mode induces important nonanalytic behaviour in hadronic observables
as a function of quark mass.  This makes extrapolations of hadronic
observables highly nontrivial
\cite{Leinweber:1992hj,Leinweber:1998ej,Leinweber:1999ig,%
Leinweber:2000sa,Hackett-Jones:2000qk,Hackett-Jones:2000js,%
Leinweber:2001ui,Detmold:2001jb,Leinweber:2001ac,Detmold:2001hq,%
Young:2002cj,Wright:2002vn,Young:2002ib,Leinweber:2003dg,Cloet:2003jm}.
Significant curvature is encountered as one approaches the chiral
limit.  Indeed there is some controversy on the optimal methodology
for achieving systematic errors within the 1\% bound
\cite{Bernard:2002yk}.

The origin of significant nonanalytic behaviour lies in the dressings
of hadrons by light pseudoscalar mesons.  While many studies of chiral
nonanalytic behaviour are formulated in the infinite-volume continuum
limit, it is important to understand that the nonanalytic behaviour is
intimately linked to the finite volume of the lattice.  The momenta
available to the hadrons participating in the loop integrals which
give rise to the nonanalytic terms of the chiral expansion are
modified in a finite volume.  For $P$-wave intermediate states there
is a threshold effect \cite{Leinweber:2001ac} where there is no
strength in the momentum integral until the first nontrivial momentum
($2 \pi / L$) is reached.  Even then the number of discrete momenta
available to the integral is governed by the number of lattice sites
($L/a$).

Systematic errors in the extrapolation of the finite lattice spacing
to the continuum limit have been minimized through the development of
improved actions \cite{Luscher:1996sc,Zanotti:2001yb,overlap}
displaying excellent scaling properties
\cite{Edwards:1998nh,Leinweber:2002bw,Dong:2000mr}.  As such, understanding
the entangled properties of chiral nonanalytic behavior on a finite
volume at a quantitative level with systematic errors at the level of
1\% is a problem which remains at the forefront of lattice gauge
theory.

For the foreseeable future, extrapolations will continue to be
required to connect lattice simulation results to physical
observables.  Hence, the development of systematically accurate chiral
extrapolation techniques is of central importance to the field
\cite{Young:2002ib,Leinweber:2003dg,Donoghue:1998bs}. 

Fortunately it is possible to test the development of chiral
extrapolation techniques today.  The key point is that one can probe
the chiral regime of quenched or partially quenched QCD using fermion
actions with improved \cite{Zanotti:2001yb} or perfect \cite{overlap}
chiral symmetry properties.  Thus, one can confront the analytic
techniques of chiral effective field theory with numerical simulation
results and test the extent to which effective field theory reproduces
the exact results of numerical simulations.

This investigation will establish the leading chiral nonanalytic
behaviour of quark sector contributions to octet baryon magnetic
moments in full, quenched and partially quenched QCD.  As we will see,
significant nonanalytic behavior remains for some quark-sector
contributions to baryon magnetic moments making these observables
ideal for the confrontation of effective field theory and lattice QCD.

Separation of the valence and sea-quark-loop contributions to the
meson cloud of full QCD hadrons is a non-trivial task.  Early
calculations addressing the meson cloud of mesons employed a
diagrammatic method \cite{Sharpe:1992ft}.  The formal theory of
quenched chiral perturbation theory (Q$\chi$PT) was subsequently
established in Ref.~\cite{Bernard:1992mk}.  There, meson properties
were examined in a graded-symmetry formulation where extra commuting
ghost-quark fields are introduced to eliminate the dependence of the
path integral on the fermion-matrix determinant.  This approach was
extended to the baryon sector in Ref.~\cite{Labrenz:1996jy}.

While the graded-symmetry formalism is essential to establishing the
field theoretic properties of Q$\chi$PT, it is desirable to formulate
an efficient and perhaps more intuitive approach to the calculation of
quenched chiral coefficients.  Rather than introducing extra degrees
of freedom to remove the effects of sea-quark loops, the approach
described herein introduces a formalism for the identification and
calculation of sea-quark-loop contributions to baryon properties,
allowing the systematic separation of valence- and sea-quark
contributions to baryon form factors in general.  Upon removing the
contributions of sea-quark-loops, one arrives at the conventional view
of quenched chiral perturbation theory.

A brief account of these methods is published in
Ref.~\cite{Leinweber:2001jc}.  Since this presentation, there has been
a resurgence in quenched and partially-quenched $\chi$PT calculations
of baryon magnetic moments \cite{Savage:2001dy,Chen:2001yi}.  In
particular, the magnetic moments of octet baryons have been examined
\cite{Savage:2001dy} using the formal approach of Q$\chi$PT
\cite{Labrenz:1996jy}.  There the leading-nonanalytic (LNA) behavior
of the magnetic moment for each baryon of the octet is calculated.
Upon summing the quark sector contributions obtained from the
diagrammatic method presented here, one finds complete agreement
between the diagrammatic and formal approaches.  The approaches are
equivalent. 

In its standard implementation, the formal approach completely
eliminates all sea-quark-loop contributions to quenched baryon
moments.  However, sea-quark loops do make a contribution to matrix
elements in the quenched approximation.  Insertion of the current in
calculating the three-point correlation function provides pair(s) of
quark-creation and annihilation operators.  These can be contracted
with the quark field operators of the hadron interpolating fields
providing ``connected insertions'' of the current, or self-contracted
to form a direct sea-quark-loop contribution or ``disconnected
insertion'' of the current.  The latter contributions to baryon
electromagnetic form factors are under intense investigation in
quenched simulations
\cite{Dong:1997xr,Mathur:2000cf,Wilcox:2000qa,Lewis:2002ix}.  Hence in
formulating quenched chiral perturbation theory it is important to
provide an opportunity to include these particular sea-quark-loop
contributions.  A more flexible approach to the calculation of
quenched chiral coefficients is desirable.

Moreover, the present calculations
\cite{Savage:2001dy,Chen:2001yi,Jenkins:1992pi} of chiral nonanalytic
behavior in baryon magnetic moments focus on bulk baryon properties.
The formalism presented here provides a method for the isolation of
individual quark sector contributions
\cite{Leinweber:1990dv,Leinweber:1992hy,Leinweber:1992pv,%
Leinweber:1993nr,Leinweber:1995ie,Leinweber:1999nf} to form factors in
full, quenched and partially-quenched QCD.  Individual quark sector
contributions to the nucleon magnetic moments are under experimental
investigation \cite{Hasty:2001ep,Aniol:2000at,A4G0} where the
strange-quark contribution to the nucleon moment is of paramount
interest.  Understanding the manner in which quarks compose baryons is
essential to a complete understanding of QCD.

In the process, we will see that it is possible to separate valence-
and sea-quark contributions to baryon form factors in {\it full} QCD.
This separation is of significant value as contributions from
disconnected insertions of the current are difficult to determine in
lattice QCD simulations.  Moreover, chiral nonanalytic behaviour of
quark sector contributions can be significantly enhanced when
direct sea-quark loop couplings to the current are removed.

In contrast to conventional calculations of chiral nonanalytic
structure, $SU(3)$-flavor violations in both meson {\it and} baryon
masses are accounted for in the following.  These are particularly
important for $K$-meson dressings of hyperons where the baryon mass
splitting can be positive or negative, suppressing or enhancing
contributions depending on whether the intermediate baryon is heavier
or lighter respectively.  Incorporation of the baryon mass splittings
significantly alters the functional structure and associated curvature
of the nonanalytic terms.  $SU(3)$-flavor violations will also affect
the analytic terms of the chiral expansion.  The leading constant, the
coefficient of $m_\pi^2\ (\propto m_q)$ etc.\ will all exhibit a
strangeness dependence due to symmetry breaking.  Ultimately, these
coefficients will be determined by matching the chiral expansion to
lattice QCD simulation results \cite{Young:2002ib,Leinweber:2003dg}.

In principle, the axial couplings, $D$ and $F$, are defined and
specified in the chiral $SU(3)$-flavor symmetric limit.  The mass
dependence of the couplings is incorporated in the chiral expansion
through higher-order terms.  Upon formulating the chiral expansion in
the $SU(2)$ limit with explicit and substantial $SU(3)$-flavor
symmetry breaking as is physically realized, one can break the
symmetry of the axial couplings.  However, previous examinations of
the chiral corrections to baryon axial currents suggest that such
phenomenological flavor-symmetry breaking is small
\cite{Jenkins:1991es}.

Similarly, $SU(3)$ flavor-symmetry breaking is often incorporated by
allowing the decay constant of the kaon to exceed that of the pion,
$f_K \simeq 1.2\, f_\pi$.  In the past, $K$-loop contributions were
often found to be uncomfortably large when matching to phenomenology
and the 30\% reduction obtained in using $f_K$ in place of $f_\pi$
proved to be helpful \cite{Jenkins:1992pi}.  However, it is not yet
clear whether this phenomenological suppression of $K$-loops through
the use of $f_K$ is necessary when regulators suppressing short
distance physics are used in chiral effective field theory.

To make the numerical results of this study readily accessible and of
the widest utility, the established tree-level axial couplings $F =
0.50$ and $D = 0.76$ are adopted with the $SU(3)$-symmetric limit of
$f_\pi = f_K = 93$ MeV for the meson decay constants.

%\begin{equation}
%\mu = \mu_0 + \chi_{\eta'}\, \log \left ( \frac{m_\pi^2}{\Lambda^2} 
%+ \chi_{\pi}\, m_\pi + \chi_{K}\, m_K 
%+ \mu_2 \, m_\pi^2 + 
%\end{equation}

In summary, the purpose of this study is:
\begin{enumerate}
\item To provide the first calculation of the leading nonanalytic
  behavior of quark-flavor contributions to baryon magnetic moments
  in full QCD,
\item quenched QCD, and
\item partially-quenched QCD.
\item To separate valence- and sea-quark contributions to baryon form
  factors in {\it full} QCD at the individual quark-flavor level.
\item To extend previous baryon-moment calculations
  \cite{Savage:2001dy,Chen:2001yi,Jenkins:1992pi} to include both
  meson and baryon mass splittings in $SU(3)$-flavor violations, as is
  done for items 1 through 4 above.
\item To identify quark-flavor channels displaying significant chiral
  nonanalytic behavior in the quenched approximation or revealing the
  artifacts of the quenched approximation.
\item To establish the diagrammatic method for calculating the chiral
  coefficients of quenched and partially-quenched QCD in the baryon
  sector.  The method is rapid, intuitive and transparent, allowing
  complete flexibility in the consideration of quark contributions to
  baryon form factors.
\end{enumerate}
It will become apparent that this technique and most of the results
may be applied to other baryon form factor studies in general.

Sec.~\ref{sec:mass} presents the essential concepts for isolating and
calculating sea-quark loop contributions to baryon properties and
proves the technique via a consideration of baryon masses.
The derivation of the quenched chiral coefficients for the
quark-sector contributions to the quenched magnetic moments of octet
baryons is described in Sec.~\ref{sec:mom}.  The technique provides a
separation of magnetic moment contributions into ``total'' full-QCD
contributions, ``direct sea-quark loop'' and ``valence''
contributions of full-QCD.  The latter are obtained by removing the
direct-current coupling to sea-quark loops from the total
contributions.  Upon further removing ``indirect sea-quark loop''
contributions, one obtains the ``quenched valence'' contributions, the
conventional view of the quenched approximation.  We will use the
quoted terms for reference to these contributions in the following.
Sec.~\ref{sec:mom} also accounts for both baryon mass and meson mass
violations of $SU(3)$-flavor symmetry.  The quenched $\eta'$ gives rise
to new nonanalytic behavior \cite{Savage:2001dy} and this is briefly
reviewed in Sec.~\ref{subsec:eta}.

A comprehensive examination of the individual quark-sector
contributions to octet baryon magnetic moments is presented in
Sec.~\ref{sec:results}.  General expressions are accompanied by
numerical evaluations to identify channels of particular interest.
Partially-quenched results are presented in Sec.~\ref{sec:partQuench}.
Sec.~\ref{sec:summary} provides a summary of the highlights of the
findings.

\end{section}

%%%%%%%%%%%%%%%%%%%%%%%%%%%%%%%%%%%%%%%%%%%%%%%%%%%%%%%%%%%%%%%%%%%%%%%%%%%

\begin{section}{QUENCHED BARYON MASSES}
\label{sec:mass}

\begin{subsection}{Formalism}

The $SU(3)$-flavor invariant couplings are described in the standard
notation by defining
\begin{equation}
  B = \left( \begin{array}{ccc}
      {\displaystyle\frac{\Sigma^{0}}{\sqrt{2}}+\frac{\Lambda}{\sqrt{6}}}
               &  \Sigma^{+}  &  p  \\[2mm]
      \Sigma^{-} & {\displaystyle-\frac{\Sigma^{0}}{\sqrt{2}}
                   +\frac{\Lambda}{\sqrt{6}}}  &  n \\[2mm]
      \Xi^{-} & \Xi^{0} &  {\displaystyle-\frac{2\Lambda}{\sqrt{6}}}
             \end{array} \right) \, ,
\end{equation}
\begin{equation}
   P_{\rm oct} = \left( \begin{array}{ccc}
      {\displaystyle\frac{\pi^{0}}{\sqrt{2}}+\frac{\eta}{\sqrt{6}}}
             & \pi^{+}  &  K^{+}  \\[2mm]
      \pi^{-} & {\displaystyle-\frac{\pi^{0}}{\sqrt{2}}
         +\frac{\eta}{\sqrt{6}}}  &   K^{0} \\[2mm]
      K^{-}  &  \overline{{K}^{0}}
             &  {\displaystyle-\frac{2\eta}{\sqrt{6}}}
             \end{array} \right) \, ,
\end{equation}
and
\begin{equation}
  P_{\rm sin} = {1 \over \sqrt{3}} \, {\rm diag} ( \eta^\prime, \eta^\prime,
  \eta^\prime )\, .
\end{equation}
The $SU(3)$-invariant combinations are
\begin{eqnarray}
  \left[\overline{B}BP\right]_{F} 
&=& {\rm Tr}(\overline{B}P_{\rm oct}B)-{\rm Tr}(\overline{B}BP_{\rm oct}),\\
  \left[\overline{B}BP\right]_{D} 
&=& {\rm Tr}(\overline{B}P_{\rm oct}B)+{\rm Tr}(\overline{B}BP_{\rm oct}),\\
  \left[\overline{B}BP\right]_{S} 
  &=& {\rm Tr}(\overline{B}B){\rm Tr}(P_{\rm sin}) \, .
\end{eqnarray}
The following calculations are simplified through the use of the
corresponding interaction Lagrangians \cite{Rijken:1998yy}.  The octet
interaction Lagrangian is
\begin{eqnarray}
   {\cal L}_{\rm int}^{\rm oct} &=&
  -f_{N\!N\pi}(\overline{N}{\tau^{\rm T}}N)\!\cdot\!{\pi}
  +if_{\Sigma\Sigma\pi}(\overline{{\Sigma}}\!\times\!{\Sigma})
      \!\cdot\!{\pi} \nonumber \\
 &&-f_{\Lambda\Sigma\pi}(\overline{\Lambda}{\Sigma}+
      \overline{{\Sigma}}\Lambda)\!\cdot\!{\pi} 
   -f_{\Xi\Xi\pi}(\overline{\Xi}{\tau^{\rm T}}\Xi)\!\cdot\!{\pi} \nonumber \\
 &&-f_{\Lambda N\!K}\left[(\overline{N}K)\Lambda
         +\overline{\Lambda}(\overline{K}N)\right] \nonumber \\
 &&-f_{\Xi\Lambda K}\left[(\overline{\Xi}K_{c})\Lambda
         +\overline{\Lambda}(\overline{K_{c}}\Xi)\right] \nonumber \\
 &&-f_{\Sigma N\!K}\left[\overline{{\Sigma}}\!\cdot\!
         (\overline{K}{\tau^{\rm T}}N)+(\overline{N}{\tau^{\rm T}}K)
         \!\cdot\!{\Sigma}\right] \nonumber \\
 &&-f_{\Xi\Sigma K}\left[\overline{{\Sigma}}\!\cdot\!
       (\overline{K_{c}}{\tau^{\rm T}}\Xi)
     +(\overline{\Xi}{\tau^{\rm T}}K_{c})\!\cdot\!{\Sigma}\right]
                                         \nonumber\nonumber \\
 &&-f_{N\!N\eta}(\overline{N}N)\eta
   -f_{\Lambda\Lambda\eta}(\overline{\Lambda}\Lambda)\eta \nonumber \\
 &&-f_{\Sigma\Sigma\eta}(\overline{{\Sigma}}\!\cdot\!
       {\Sigma})\eta
   -f_{\Xi\Xi\eta}(\overline{\Xi}\Xi)\eta. 
\end{eqnarray}
and the singlet interaction Lagrangian is
\begin{eqnarray}
   {\cal L}_{\rm int}^{\rm sin} &=&
       -f_{N\!N\eta'}(\overline{N}N)\eta'
       -f_{\Lambda\Lambda\eta'}(\overline{\Lambda}\Lambda)\eta' \nonumber \\
     &&-f_{\Sigma\Sigma\eta'}(\overline{{\Sigma}}\!\cdot\!
        {\Sigma})\eta'
       -f_{\Xi\Xi\eta'}(\overline{\Xi}\Xi)\eta',
\end{eqnarray}
where
\begin{equation}
  \begin{array}{ll}
  N=\left(\begin{array}{c} p \\ n \end{array} \right) \, , \quad
  &\Xi=\left(\begin{array}{c} \Xi^{0} \\ \Xi^{-} \end{array} \right)
  \, , \nonumber \\
  K=\left(\begin{array}{c} K^{+} \\ K^{0} \end{array} \right) \, , \quad
  &K_{c}=\left(\begin{array}{c} \overline{K^{0}} \\
               -K^{-} \end{array} \right) \, . \nonumber \\
  \end{array}
\end{equation}
The octet meson-baryon couplings are expressed in terms of the $F$ and
$D$ coupling coefficients as follows:
\begin{equation}
  \begin{array}{lll}
   f_{NN\pi}                 = F + D,                           \ \ \  &
   f_{\Lambda NK}            =-\frac{1}{\sqrt{3}}\, (3 F + D),  \\
   f_{NN\eta}            = \frac{1}{\sqrt{3}}\, (3 F - D),  \ \ \  &
   f_{\Sigma\Sigma\pi}       = 2\, F,                           \\
   f_{\Xi\Lambda K}          = \frac{1}{\sqrt{3}}\, (3 F - D),  \ \ \  &
   f_{\Lambda\Lambda\eta}=-\frac{2}{\sqrt{3}}\, D,          \\
   f_{\Lambda\Sigma\pi}      = \frac{2}{\sqrt{3}}\, D,          \ \ \  &
   f_{\Sigma NK}             = D - F,                           \\
   f_{\Sigma\Sigma\eta}  = \frac{2}{\sqrt{3}}\, D,          \ \ \  &
   f_{\Xi\Xi\pi}             = F - D,                           \\
   f_{\Xi\Sigma K}           =-(F + D),                         \ \ \  &
   f_{\Xi\Xi\eta}        =-\frac{1}{\sqrt{3}}\, (3 F + D),
  \end{array}
\label{goct}
\end{equation}
and the singlet couplings satisfy
\begin{equation}
   f_{NN\eta^\prime} = f_{\Lambda\Lambda \eta^\prime}
    = f_{\Sigma\Sigma\eta^\prime} = f_{\Xi\Xi \eta^\prime} \, .
\label{gsin}
\end{equation}
The light quark content of
\begin{equation}
\mid\! \eta^\prime \,\rangle = {1 \over \sqrt{3}} \left (
                          \mid\! \overline u u \,\rangle 
                        + \mid\! \overline d d \,\rangle 
                        + \mid\! \overline s s \,\rangle 
                        \right ) \, ,
\end{equation}
and
\begin{equation}
\mid\! \eta \,\rangle = {1 \over \sqrt{6}} \left (
                          \mid\! \overline u u \,\rangle 
                        + \mid\! \overline d d \,\rangle 
                        - 2\, \mid\! \overline s s \,\rangle 
                        \right ) \, ,
\end{equation}
mesons suggests 
\begin{equation}
f_{NN\eta^\prime} = \sqrt{2} f_{NN\eta} \, .
\end{equation}
This relation between nucleon octet and singlet couplings is commonly
used in Q$\chi$PT calculations to estimate $\eta^\prime$ couplings to
octet baryons.  In the following, numerical estimates are based on the
tree-level axial couplings $F = 0.50$ and $D = 0.76$ with $f_\pi = f_K
= 93$ MeV.

   The leading-nonanalytic term of the chiral expansion provides the
dominant source of rapid variation in baryon magnetic moments.
Taking the $N$-$\Delta$ mass splitting to be of zeroth chiral
order, the $\Delta$ resonance provides the next-to-leading nonanalytic
(NLNA) contribution.  This contribution makes only a small correction
to the proton magnetic moment obtained from the chiral extrapolation
of lattice QCD simulation results.  The $N$-$\Delta$ mass splitting
suppresses the nonanalytic curvature, and when combined with the
analytic terms of the chiral expansion, only a small enhancement of
the proton moment is observed.  As the NLNA $\Delta$ contribution is
added, the proton moment increases from 2.61 to 2.66 $\mu_N$
\cite{Cloet:2003jm}.  This 2\% effect at the physical point is
unlikely to be observed in lattice QCD simulation results for some
time.  As such, we do not encumber the reader with these small
contributions.

\end{subsection}

\begin{subsection}{Baryon Mass}

\begin{figure*}[t]
% \begin{figure*}[t]
% \begin{center}
% \setlength{\unitlength}{1.6cm}
% \setlength{\fboxsep}{0cm}
% \begin{picture}(10.5,7.9)
% \put(1.55,3.70){\begin{picture}(3,4.5)\put(0,0){
% \epsfig{file=ProtonPi0.ps,width=4.8cm}}\end{picture}}
% \put(5.95,4.67){\begin{picture}(3,4.5)\put(0,0){
% \epsfig{file=ProtonPiP.ps,width=4.8cm}}\end{picture}}
% \put(1.55,0.00){\begin{picture}(3,4.5)\put(0,0){
% \epsfig{file=ProtonPiM.ps,width=4.8cm}}\end{picture}}
% \put(5.95,0.00){\begin{picture}(3,4.5)\put(0,0){
% \epsfig{file=ProtonK.ps,  width=4.8cm}}\end{picture}}
% \end{picture}
% \end{center}
\begin{center}
\setlength{\unitlength}{1.6cm}
\setlength{\fboxsep}{0cm}
\begin{picture}(10.5,6.6)
\put(0.0,2.53){\begin{picture}(3,4.5)\put(0,0){
\epsfig{file=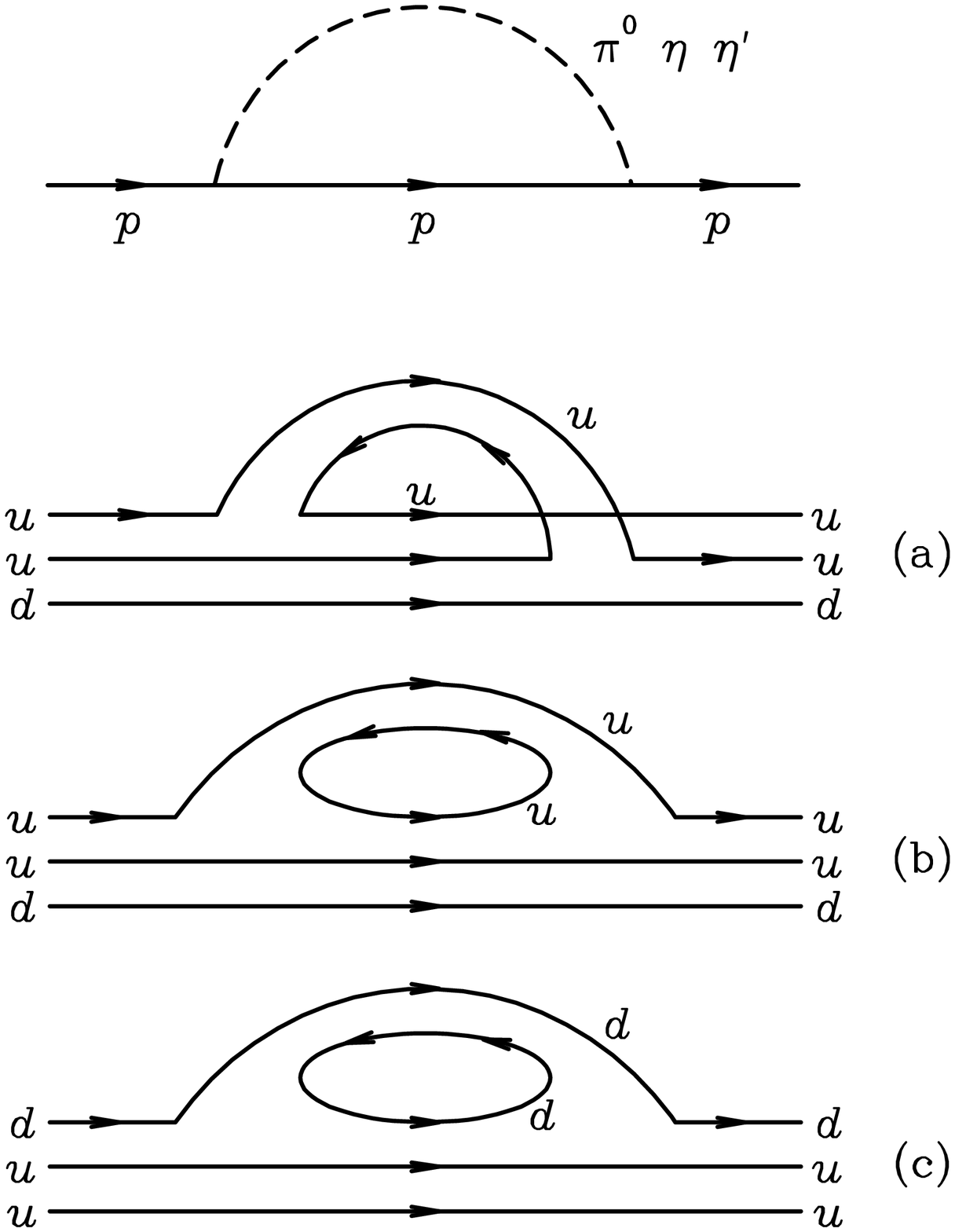,width=4.8cm}}\end{picture}}
\put(3.7,3.5){\begin{picture}(3,4.5)\put(0,0){
\epsfig{file=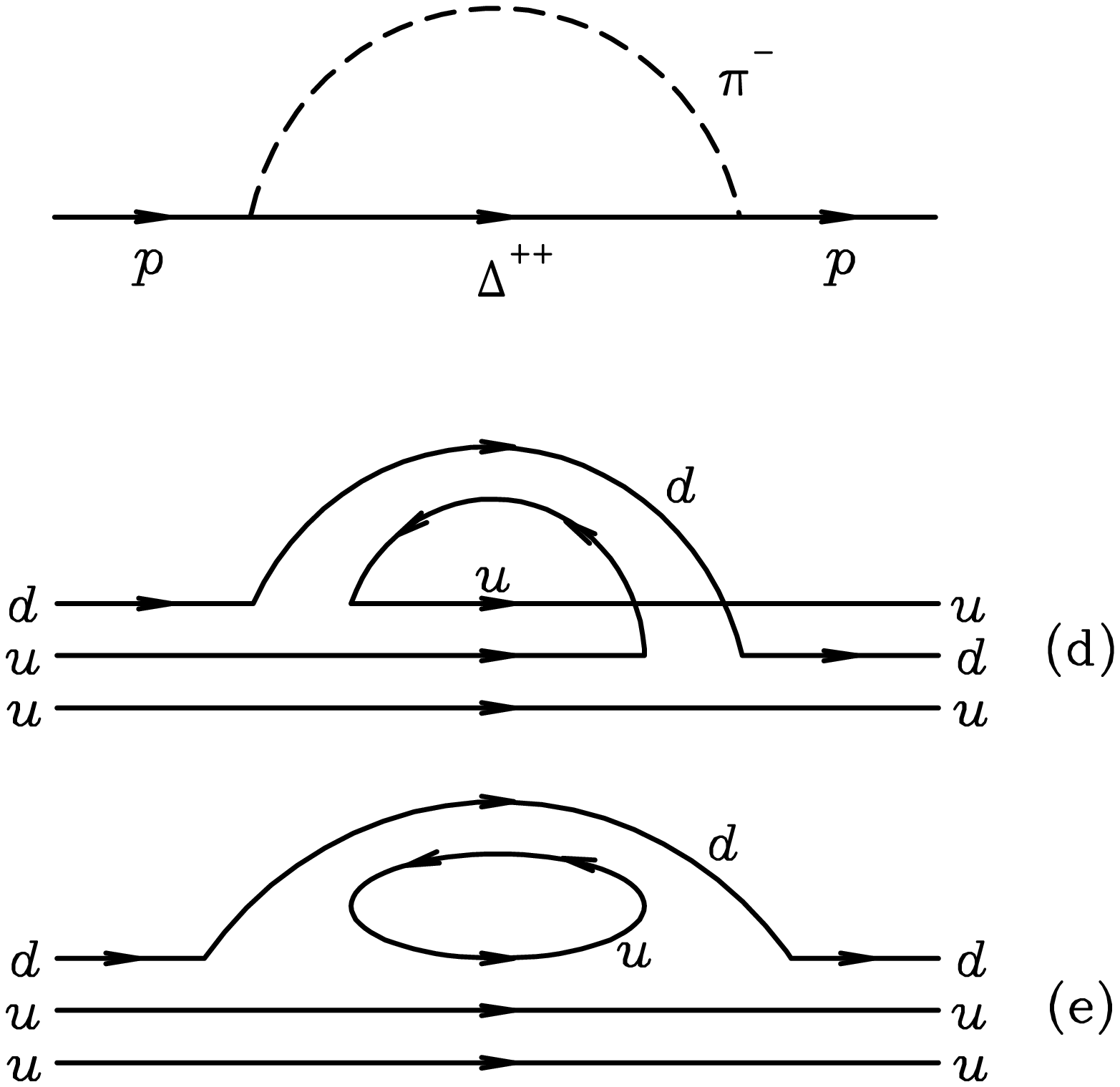,width=4.8cm}}\end{picture}}
\put(3.7,0){\begin{picture}(3,4.5)\put(0,0){
\epsfig{file=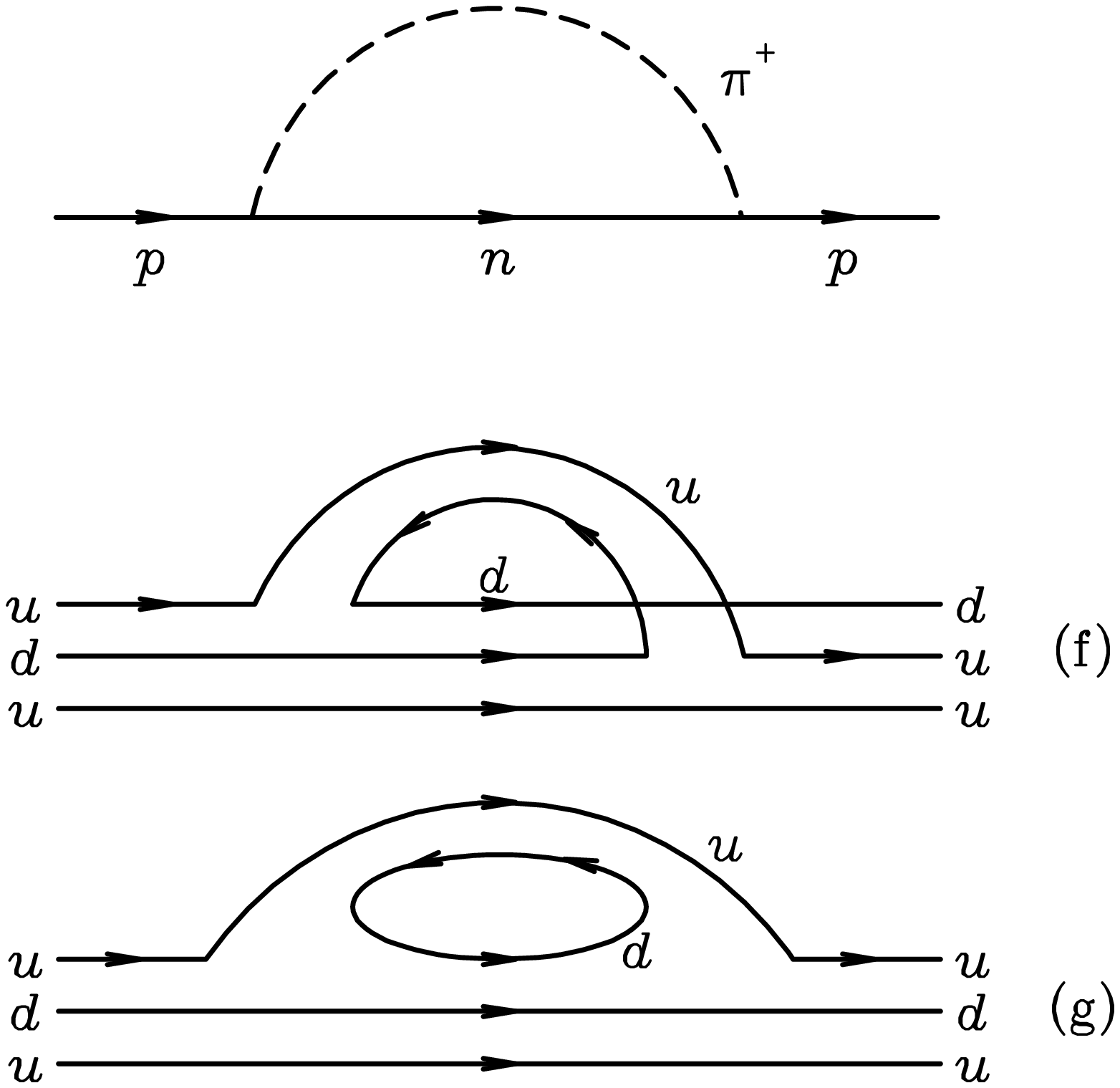,width=4.8cm}}\end{picture}}
\put(7.4,0){\begin{picture}(3,4.5)\put(0,0){
\epsfig{file=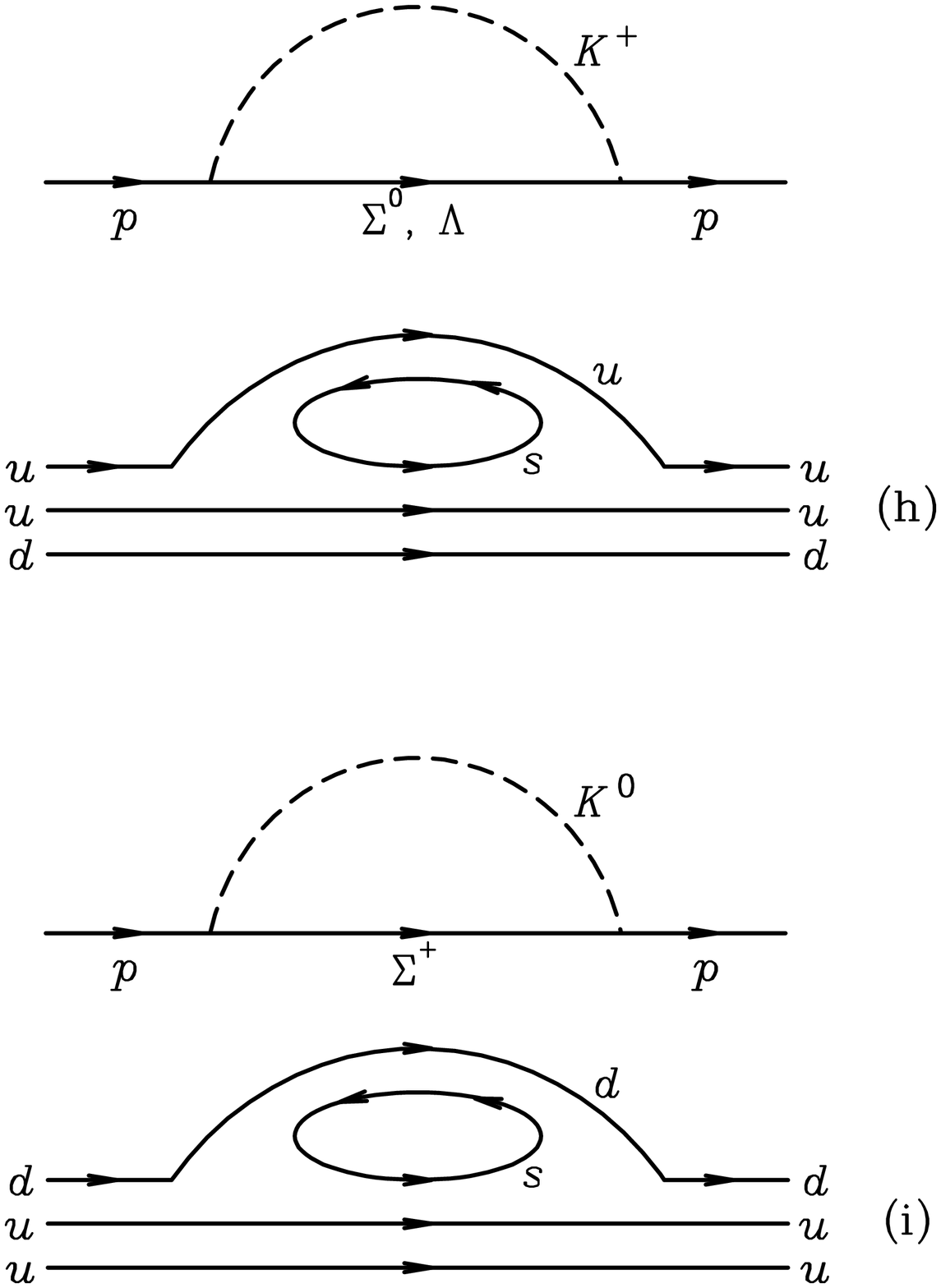,width=4.8cm}}\end{picture}}
\end{picture}
\end{center}
\caption{The pseudo-Goldstone meson cloud of the proton and
associated quark flow diagrams.}
\label{ProtonCloud}
\end{figure*}

To calculate the quenched chiral coefficients, we begin by calculating
the total full-QCD contribution in the limit where the $\eta$ and
$\eta'$ mesons are taken to be degenerate with the pion.  In the
quenched approximation, quark loops which otherwise break this
degeneracy are absent.  The quark flow diagrams of
Fig.~\ref{ProtonCloud} illustrate the processes which give rise to the
LNA behavior of proton observables.  Table~\ref{tab:massCloud}
summarizes the contributions of the $\pi$-, $\eta$-, $\eta'$- and
$K$-cloud diagrams of Fig.~\ref{ProtonCloud} labeled by the
corresponding quark-flow diagrams.  Summation of these couplings and
incorporation of the factors from the loop integral provides the LNA
term proportional to $m_\pi^3$ of
\begin{equation}
- ( 3 F^2 + D^2 ) \, {m_\pi^3 \over 8 \pi  f_\pi^2} \, .
\end{equation}

\begin{table}[t]
\caption{Meson-cloud contributions of Fig.~\protect\ref{ProtonCloud}
  in full QCD.  $\eta$ and $\eta'$ masses are set degenerate with the
  pion in anticipation of quenching the theory.}
\label{tab:massCloud}
\begin{ruledtabular}
\begin{tabular}{ccccc}
\noalign{\smallskip}
Fig.    &Channel          &Mass        &Coupling            &Coupling \\
\hline		      
\noalign{\smallskip}
a,b,c   &$p\, \pi^0$      &$N \pi$     &$f_{NN\pi}^2$       &$(F+D)^2$ \\
a,b,c   &$p\, \eta$       &$N \pi$     &$f_{NN\eta}^2$      &$(3F-D)^2/3$ \\
a,b,c   &$p\, \eta'$      &$N \pi$     &$f_{NN\eta'}^2$     &$2(3F-D)^2/3$ \\
f,g     &$n\, \pi^+$      &$N \pi$     &$2f_{NN\pi}^2$      &$2(F+D)^2$ \\
h       &$\Lambda\,  K^+$ &$\Lambda K$ &$f_{\Lambda N K}^2$ &$(3F+D)^2/3$ \\
h       &$\Sigma^0\, K^+$ &$\Sigma K$  &$f_{\Sigma N K}^2$  &$(D-F)^2$ \\
i       &$\Sigma^+\, K^0$ &$\Sigma K$  &$2f_{\Sigma N K}^2$ &$2(D-F)^2$ \\
\end{tabular}
\end{ruledtabular}

\vspace{18pt}

\caption{Sea-quark-loop contributions of Fig.~\protect\ref{ProtonCloud}.}
\label{tab:massLoop}
\begin{ruledtabular}
\begin{tabular}{ccccc}
\noalign{\smallskip}
Fig.    &Channel          &Mass        &Coupling            &Coupling \\
\hline			  
\noalign{\smallskip}
b       &$\Lambda\,  K^+$ &$N \pi$     &$f_{\Lambda N K}^2$ &$(3F+D)^2/3$ \\
b       &$\Sigma^0\, K^+$ &$N \pi$     &$f_{\Sigma N K}^2$  &$(D-F)^2$ \\
c       &$\Sigma^+\, K^0$ &$N \pi$     &$2f_{\Sigma N K}^2$ &$2(D-F)^2$ \\
e       &$\Sigma^+\, K^0$ &$N \pi$     &$2f_{\Sigma N K}^2$ &$2(D-F)^2$ \\
g       &$\Lambda\,  K^+$ &$N \pi$     &$f_{\Lambda N K}^2$ &$(3F+D)^2/3$ \\
g       &$\Sigma^0\, K^+$ &$N \pi$     &$f_{\Sigma N K}^2$  &$(D-F)^2$ \\
h       &$\Lambda\,  K^+$ &$\Lambda K$ &$f_{\Lambda N K}^2$ &$(3F+D)^2/3$ \\
h       &$\Sigma^0\, K^+$ &$\Sigma K$  &$f_{\Sigma N K}^2$  &$(D-F)^2$ \\
i       &$\Sigma^+\, K^0$ &$\Sigma K$  &$2f_{\Sigma N K}^2$ &$2(D-F)^2$ \\
\end{tabular}
\end{ruledtabular}
\end{table}

The separation of the meson cloud into valence and sea-quark
contributions is shown in the quark-flow diagrams labeled by letters.
To quench the theory, one must understand the chiral behavior of the
valence-quark loops of Figs.~\ref{ProtonCloud}(a), (d) and (f) and the
sea-quark loops of Figs.~\ref{ProtonCloud}(b), (c), (e), (g), (h) and
(i) separately.  If one can isolate the behavior of the diagrams
involving a quark loop, then one can use the known LNA behavior of the
full meson-based diagrams to extract the corresponding valence-loop
contributions.

For example, Fig.~\ref{ProtonCloud}(b) involves a $u$-quark loop where
no exchange term is possible.  Thus the $u$-quark in the loop is
distinguishable from all the other quarks in the diagram.  The chiral
structure of this diagram is therefore identical to that for a
``strange'' quark loop, as illustrated in Fig.\ \ref{ProtonCloud}(h),
$(p \to K^+ \Sigma^0\ {\rm or}\ p \to K^+ \Lambda )$ provided the
``strange'' quark in {\it this} case is understood to have the same
mass as the $u$-quark.

Similarly, the corresponding hadron diagram which gives rise to the
LNA structure of Fig.\ \ref{ProtonCloud}(c) is therefore the
$K^0$-loop diagram of Fig.~\ref{ProtonCloud}(i), with the
distinguishable ``strange'' quark mass set equal to the mass of the
$d$-quark.  That is, the intermediate $\Sigma$-baryon mass appearing in the
$K^0$-loop diagram is degenerate with the nucleon.  Similarly, the
``kaon'' mass is degenerate with the pion.

The sum of the first two lines of Table~\ref{tab:massLoop} provides
the contribution of diagram Fig.~\ref{ProtonCloud}(b).  The third line
provides the contribution of Fig.~\ref{ProtonCloud}(c).  Similar
arguments allow one to establish the remaining loop contributions to
the light-meson cloud.  Summing the couplings of
Table~\ref{tab:massLoop} indicates sea-quark-loops contribute a term
\begin{equation}
- ( 9 F^2 - 6 F D + 5 D^2 ) \, {m_\pi^3 \over 24 \pi
f_\pi^2}  \, ,
\end{equation}
to the LNA behavior of the nucleon such that the net quenched
contribution proportional to $m_\pi^3$ is
\begin{equation}
- (3 F D - D^2 ) \, {m_\pi^3 \over 12 \pi  f_\pi^2} \, ,
\end{equation}
in agreement with the formal approach of Labrenz and Sharpe
\cite{Labrenz:1996jy}.  We note that the kaon-cloud contributions of
Fig.~\ref{ProtonCloud}(h) and (i) are pure sea contributions and
trivially vanish in subtracting sea-contributions from total
contributions as outlined above.

\end{subsection}

\end{section}

%%%%%%%%%%%%%%%%%%%%%%%%%%%%%%%%%%%%%%%%%%%%%%%%%%%%%%%%%%%%%%%%%%%%%%%%%%%

\begin{section}{BARYON MAGNETIC MOMENTS}
\label{sec:mom}

\begin{subsection}{Quark-Sector Contributions to the Proton}
\label{subsec:proton}

\begin{table*}[t]
\caption{Determination of the total $u$-quark contribution to the proton
magnetic moment as illustrated in Fig.~\protect\ref{ProtonCloud}.}
\label{tab:uProtonTotal}
%\begin{sidewaystable}
%\begin{center}
\begin{ruledtabular}
\begin{tabular}{ccccccc}
Diagram   &Channel         &Mass        &Charge &Term                                      &$\beta$       &$\chi$  \\
\hline
f,g     &$n\, \pi^+$      &$N \pi$     &$+1$   &$+2 f_{NN\pi}^2       \, m_\pi$           &$-(F+D)^2$    &$-6.87$ \\
h       &$\Sigma^0\, K^+$ &$\Sigma K$  &$+1$   &$+  f_{\Sigma N K}^2  \, m_{N \Sigma K}$  &$-(D-F)^2/2$  &$-0.15$ \\
h       &$\Lambda \, K^+$ &$\Lambda K$ &$+1$   &$+  f_{\Lambda N K}^2 \, m_{N \Lambda K}$ &$-(3F+D)^2/6$ &$-3.68$ \\
\end{tabular}
\end{ruledtabular}
%\end{sidewaystable}
%\end{center}

\vspace{18pt}

\caption{Determination of direct $u$-quark sea-quark loop
contributions to the proton magnetic moment as illustrated in
Fig.~\protect\ref{ProtonCloud}.}
\label{tab:uProtonSea}
%\begin{sidewaystable}
%\begin{center}
\begin{ruledtabular}
\begin{tabular}{ccccccc}
Diagram   &Channel         &Mass     &Charge   &Term                            &$\beta$    &$\chi$  \\
\hline
b       &$\Lambda\,  K^+$ &$N \pi$  &$-1$     &$-f_{\Lambda N K}^2 \, m_\pi$   &$(3F+D)^2/6$ &$+3.68$ \\
b       &$\Sigma^0\, K^+$ &$N \pi$  &$-1$     &$-f_{\Sigma N K}^2  \, m_\pi$   &$(D-F)^2/2$  &$+0.15$ \\
e       &$\Sigma^+\, K^0$ &$N \pi$  &$-1$     &$-2f_{\Sigma N K}^2 \, m_\pi$   &$(D-F)^2$    &$+0.29$ \\
Total   &                 &         &         &                                &             &$+4.12$ \\
\end{tabular}
\end{ruledtabular}
%\end{sidewaystable}
%\end{center}

\vspace{18pt}

\caption{Indirect sea-quark loop contributions from $u$ valence quarks
to the proton magnetic moment.  Here, the $u$-valence quark forms a meson
composed with a sea-quark loop as illustrated in
Fig.~\protect\ref{ProtonCloud}.}
\label{tab:uProtonQuench}
%\begin{sidewaystable}
%\begin{center}
\begin{ruledtabular}
\begin{tabular}{ccccccc}
Diagram   &Channel         &Mass     &Charge   &Term                            &$\beta$       &$\chi$  \\
\hline 
b       &$\Lambda\,  K^+$ &$N \pi$  &$+1$     &$+f_{\Lambda N K}^2 \, m_\pi$   &$-(3F+D)^2/6$ &$-3.68$ \\
b       &$\Sigma^0\, K^+$ &$N \pi$  &$+1$     &$+f_{\Sigma N K}^2  \, m_\pi$   &$-(D-F)^2/2$  &$-0.15$ \\
g       &$\Lambda\,  K^+$ &$N \pi$  &$+1$     &$+f_{\Lambda N K}^2 \, m_\pi$   &$-(3F+D)^2/6$ &$-3.68$ \\
g       &$\Sigma^0\, K^+$ &$N \pi$  &$+1$     &$+f_{\Sigma N K}^2  \, m_\pi$   &$-(D-F)^2/2$  &$-0.15$ \\
h       &$\Sigma^0\, K^+$ &$\Sigma K$  &$+1$   &$+  f_{\Sigma N K}^2  \, m_{N \Sigma K}$  &$-(D-F)^2/2$  &$-0.15$ \\
h       &$\Lambda \, K^+$ &$\Lambda K$ &$+1$   &$+  f_{\Lambda N K}^2 \, m_{N \Lambda K}$ &$-(3F+D)^2/6$ &$-3.68$ \\
\end{tabular}
\end{ruledtabular}
%\end{sidewaystable}
%\end{center}
\end{table*}

The LNA contribution to baryon magnetic moments proportional to
$m_\pi$ or $m_K$ has its origin in couplings of the electromagnetic
(EM) current to the meson propagating in the intermediate meson-baryon
state.  In order to pick out a particular quark-flavor contribution,
one sets the electric charge for the quark of interest to one and the
charge of all other flavors to zero.

Tables~\ref{tab:uProtonTotal} through \ref{tab:uProtonQuench} report
results for the $u$-quark in the proton.  The total contributions are
calculated in the standard way, but with charge assignments for the
intermediate mesons (indicated in the Charge column) reflecting in
this case $q_u = 1$ and $q_d = q_s = 0$.  The extra baryon subscripts
on the meson masses are a reminder of the baryons participating in the
diagram to facilitate more accurate treatments of the loop
integral in which baryon mass splittings are taken into account.  The
LNA contribution is 
\begin{equation}
\beta \, {m_N \over 8 \pi f_\pi^2 } \, m_\pi
\equiv \chi \, m_\pi
\end{equation}
with $\beta$ and $\chi$ indicated in the last two columns.  Throughout
the following, the units of $\chi$ are $\mu_N/$GeV, such that when
multiplied by the pion mass in GeV, one obtains magnetic moment
contributions in units of the nuclear magneton, $\mu_N$.

The ``direct sea-quark-loop contributions'' indicated in Table
\ref{tab:uProtonSea} are contributions in which the EM current couples
to a sea-quark loop, in this case a $u$ quark.  Using the techniques
described in Sec.~\ref{sec:mass}, one can calculate the contributions
of these loops alone to the baryon magnetic moment.  The Mass column
of Table~\ref{tab:uProtonSea} is a reminder that the mass of the
``kaon'' considered in determining the coupling is actually the pion
mass for Figs.~\ref{ProtonCloud}(b) and (e).  These diagrams will
contribute, even in the quenched approximation, when disconnected
insertions of the EM current are included in simulations
\cite{Dong:1997xr,Mathur:2000cf,Wilcox:2000qa,Lewis:2002ix}.

Subtraction of these sea-quark-loop contributions from the total
contributions of \ref{tab:uProtonTotal} leaves a net valence
contribution of $-11.0\, m_\pi -0.15\, m_{N \Sigma K} -3.68\, m_{N
\Lambda K}$ in {\it full} QCD.

Table \ref{tab:uProtonQuench} focuses on diagrams in which the EM
current couples to a valence quark in a meson composed with a
sea-quark loop.  These are the ``indirect sea-quark loop''
contributions.  Subtracting off these couplings from the valence
contribution provides the net quenched valence contribution of $-3.33 \,
m_\pi$.

\begin{table*}[t]
\caption{Determination of the total $d$-quark contribution to the proton
magnetic moment as illustrated in Fig.~\protect\ref{ProtonCloud}.}
\label{tab:dProtonTotal}
%\begin{sidewaystable}
%\begin{center}
\begin{ruledtabular}
\begin{tabular}{ccccccc}
Diagram   &Channel         &Mass        &Charge &Term                                      &$\beta$       &$\chi$  \\
\hline
f,g     &$n\, \pi^+$ &$N \pi$  &$-1$     &$-2 f_{NN\pi}^2 \, m_\pi$    &$(F+D)^2$  &$+6.87$ \\
i       &$\Sigma^+\, K^0$ &$\Sigma K$  &$+1$   &$+2 f_{\Sigma N K}^2  \, m_{N \Sigma K}$  &$-(D-F)^2$  &$-0.29$ \\
\end{tabular}
\end{ruledtabular}
%\end{sidewaystable}
%\end{center}

\vspace{18pt}

\caption{Determination of direct $d$-quark sea-quark loop
contributions to the proton magnetic moment as illustrated in
Fig.~\protect\ref{ProtonCloud}.}
\label{tab:dProtonSea}
%\begin{sidewaystable}
%\begin{center}
\begin{ruledtabular}
\begin{tabular}{ccccccc}
Diagram   &Channel         &Mass     &Charge   &Term                            &$\beta$    &$\chi$  \\
\hline
c       &$\Sigma^+\, K^0$ &$N \pi$  &$-1$     &$-2f_{\Sigma N K}^2 \, m_\pi$   &$(D-F)^2$    &$+0.29$ \\
g       &$\Lambda\,  K^+$ &$N \pi$  &$-1$     &$-f_{\Lambda N K}^2 \, m_\pi$   &$(3F+D)^2/6$ &$+3.68$ \\
g       &$\Sigma^0\, K^+$ &$N \pi$  &$-1$     &$-f_{\Sigma N K}^2  \, m_\pi$   &$(D-F)^2/2$  &$+0.15$ \\
Total   &                 &         &         &                                &             &$+4.12$ \\
\end{tabular}
\end{ruledtabular}
%\end{sidewaystable}
%\end{center}

\vspace{18pt}

\caption{Indirect sea-quark loop contributions from $d$ valence quarks
to the proton magnetic moment.  Here the $d$-valence quark forms a
meson composed with a sea-quark loop as illustrated in
Fig.~\protect\ref{ProtonCloud}.}
\label{tab:dProtonQuench}
%\begin{sidewaystable}
%\begin{center}
\begin{ruledtabular}
\begin{tabular}{ccccccc}
Diagram   &Channel         &Mass     &Charge   &Term                            &$\beta$       &$\chi$  \\
\hline 
c       &$\Sigma^+\, K^0$ &$N \pi$  &$+1$     &$2f_{\Sigma N K}^2 \, m_\pi$    &$-(D-F)^2$    &$-0.29$ \\
e       &$\Sigma^+\, K^0$ &$N \pi$  &$+1$     &$2f_{\Sigma N K}^2 \, m_\pi$    &$-(D-F)^2$    &$-0.29$ \\
i       &$\Sigma^+\, K^0$ &$\Sigma K$  &$+1$   &$+2 f_{\Sigma N K}^2  \, m_{N \Sigma K}$  &$-(D-F)^2$  &$-0.29$ \\
\end{tabular}
\end{ruledtabular}
%\end{sidewaystable}
%\end{center}
\end{table*}

Tables \ref{tab:dProtonTotal} through \ref{tab:dProtonQuench} provide
a similar analysis of the $d$ quark in the proton, where $q_d = 1$ and
$q_u = q_d = 0$.  Subtraction of the direct sea-quark loop contributions of Table
\ref{tab:dProtonSea} from the total contributions of Table
\ref{tab:dProtonTotal} leaves a net valence contribution of $2.75\,
m_\pi -0.29\, m_{N \Sigma K}$.  Further removal of the indirect
sea-quark loops of Table
\ref{tab:dProtonQuench} provides the final net $d$-quark quenched
valence contribution to the proton moment of $+3.33\, m_\pi$.

\begin{table*}[t]
\caption{Determination of the total $s$-quark contribution to the proton
magnetic moment as illustrated in Fig.~\protect\ref{ProtonCloud}.  As
there are no $s$ valence quarks in the proton, the contributions are
purely sea-quark-loop contributions.}
\label{tab:sProtonTotal}
%\begin{sidewaystable}
%\begin{center}
\begin{ruledtabular}
\begin{tabular}{ccccccc}
Diagram   &Channel         &Mass        &Charge &Term                                      &$\beta$       &$\chi$  \\
\hline
h       &$\Sigma^0\, K^+$ &$\Sigma K$  &$-1$   &$-  f_{\Sigma N K}^2  \, m_{N \Sigma K}$  &$(D-F)^2/2$   &$0.15$ \\
h       &$\Lambda \, K^+$ &$\Lambda K$ &$-1$   &$-  f_{\Lambda N K}^2 \, m_{N \Lambda K}$ &$(3F+D)^2/6$  &$3.68$ \\
i       &$\Sigma^+\, K^0$ &$\Sigma K$  &$-1$   &$-2 f_{\Sigma N K}^2  \, m_{N \Sigma K}$  &$(D-F)^2  $   &$0.29$ \\
\end{tabular}
\end{ruledtabular}
%\end{sidewaystable}
%\end{center}
\end{table*}

Table \ref{tab:sProtonTotal} describes the $s$-quark contributions to
the proton magnetic moment, where $q_s = 1$ and $q_u = q_d = 0$.  As
there are no $s$ valence quarks in the proton, the contributions are
purely sea-quark-loop contributions.  The net valence contribution is
zero and there are no further quenching considerations.

Charge symmetry provides the quark-sector contributions to the neutron
magnetic moment.  For unit charge quarks, $d_n = u_p$, $u_n = d_p$
and $s_n = s_p$.

The QCD Lagrangian is flavor blind in the $SU(3)$-flavor symmetry
limit.  This independence from quark flavor is manifest in Tables
\ref{tab:uProtonSea}, \ref{tab:dProtonSea} and
\ref{tab:sProtonTotal} for the direct sea-quark loop contributions to
the proton form factor.   In each case there are three channels
for the coupling, $\beta$.  Indeed the $u$ and $d$ direct sea-quark
loop contributions are exactly equal.  However $SU(3)$-flavor
symmetry breaking due to the massive $s$ quark requires one to track
the masses of intermediate mesons and baryons, and this introduces the
$K$, $\Sigma$ and $\Lambda$ masses in Table \ref{tab:sProtonTotal}.

$SU(3)$-flavor symmetry is also manifest in the indirect sea-quark loop
contributions to the proton magnetic moment.  For example, the
$u$-quark indirect sea-quark loop result receives contributions from
each of $u$, $d$ and $s$ quark loops in Figs.~\ref{ProtonCloud}(b),
\ref{ProtonCloud}(g) and \ref{ProtonCloud}(h).  Each of these
contributions appearing in Table \ref{tab:uProtonQuench} are equal up
to symmetry breaking in the meson and baryon masses.  Similar results
hold for the $d$-valence quark of the proton in Table
\ref{tab:dProtonQuench}. 

The flavor-blind nature of QCD makes it trivial to extend this
calculation of quenched quark-sector magnetic moments to the
partially-quenched theory.  As new flavors are introduced through the
use of dynamically generated gauge fields, one simply adds the direct
and indirect sea-quark loop contributions evaluated here to the
quenched results, keeping track of the meson mass of the valence-sea
meson.  The latter is simple to do as we have already isolated each
valence quark flavor contribution to the baryon moment.  This is
described in further detail in Sec.~\ref{sec:partQuench}.

\end{subsection}

\begin{subsection}{Quark-Sector Contributions to $\Sigma^+$}
\label{subsec:sigma}

\begin{figure*}[t]
\begin{center}
\setlength{\unitlength}{1.6cm}
\setlength{\fboxsep}{0cm}
\begin{picture}(10.5,4.8)
\put(0.0,0){\begin{picture}(3,4.5)\put(0,0){
\epsfig{file=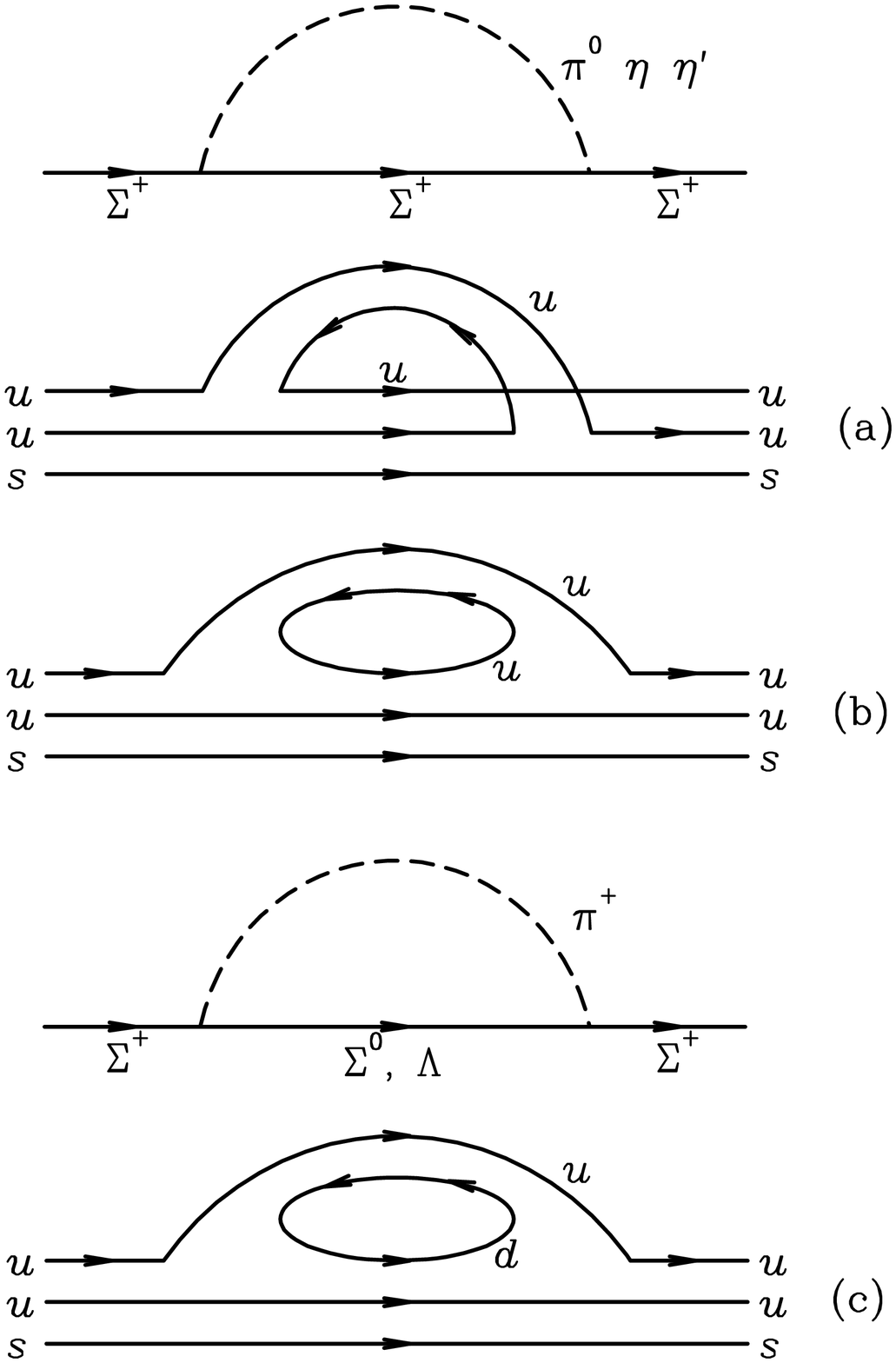,width=4.8cm}}\end{picture}}
\put(3.7,0){\begin{picture}(3,4.5)\put(0,0){
\epsfig{file=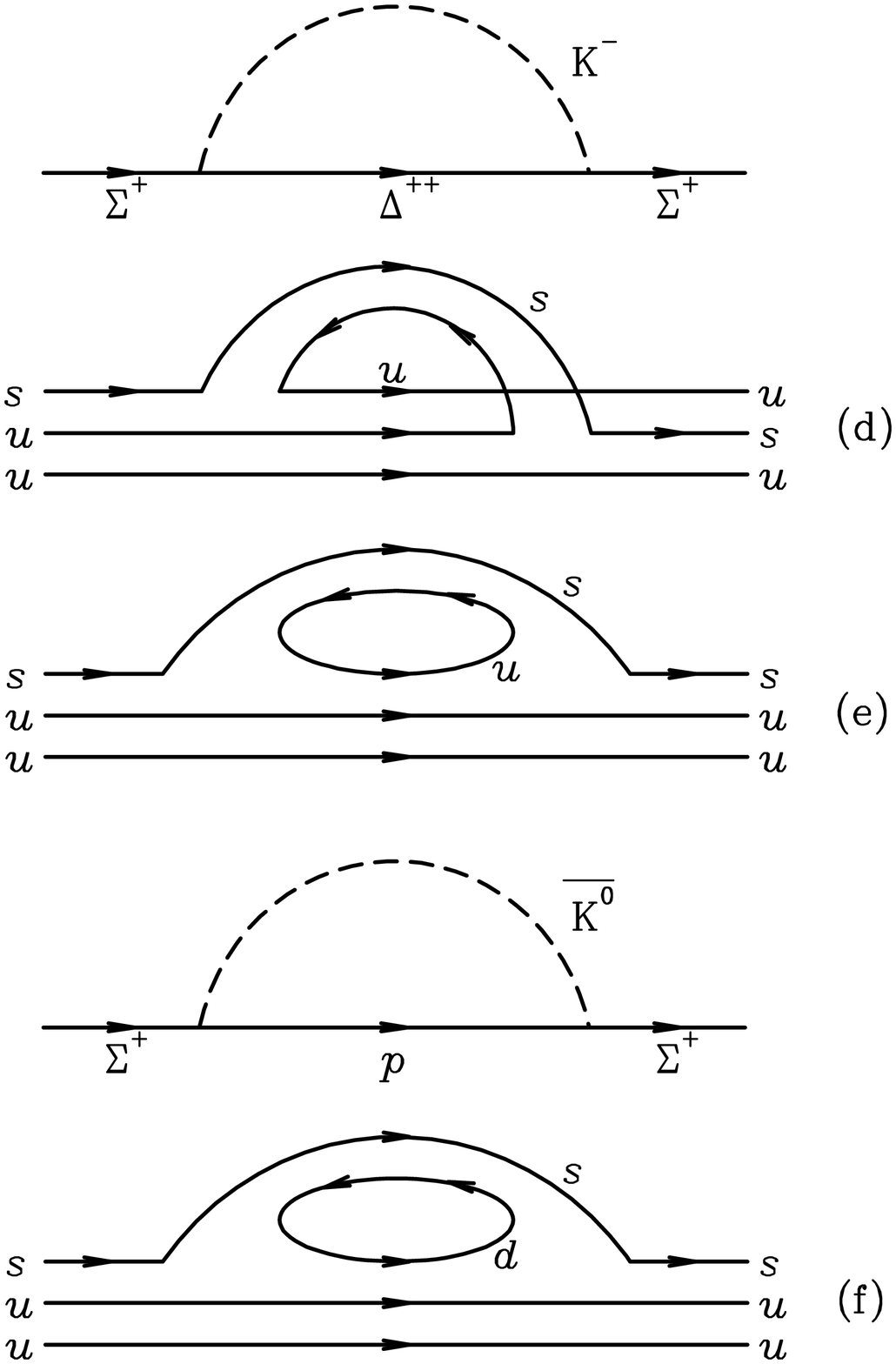,width=4.8cm}}\end{picture}}
\put(7.4,0){\begin{picture}(3,4.5)\put(0,0){
\epsfig{file=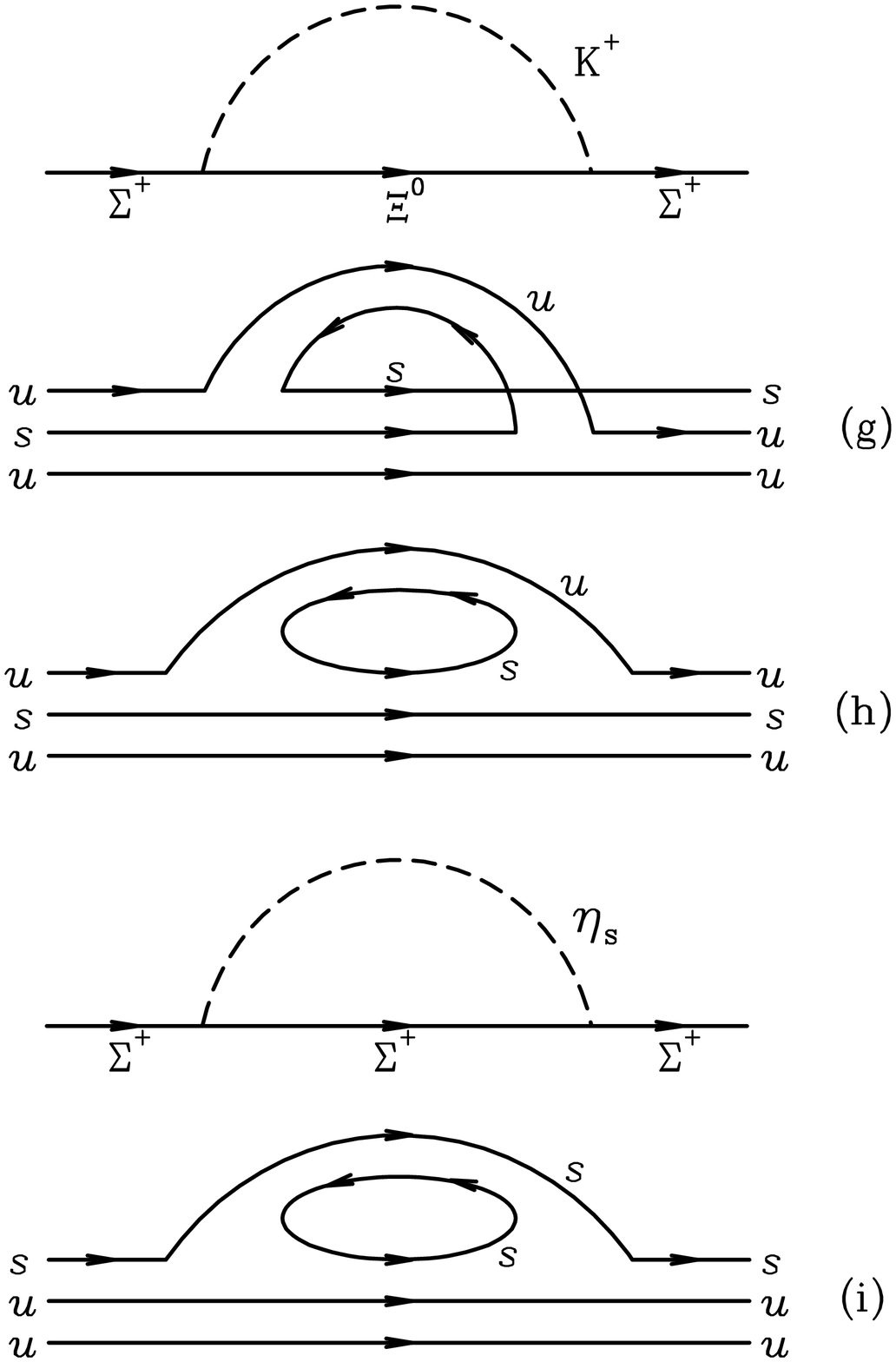,width=4.8cm}}\end{picture}}
\end{picture}
\end{center}
\caption{The pseudo-Goldstone meson cloud of $\Sigma^+$ and
associated quark flow diagrams.}
\label{SigmaCloud}
\end{figure*}

\begin{table*}[t]
\caption{Determination of the total $u$-quark contribution to the $\Sigma^+$
magnetic moment as illustrated in Fig.~\protect\ref{SigmaCloud}.}
\label{tab:uSigmaTotal}
%\begin{sidewaystable}
%\begin{center}
\begin{ruledtabular}
\begin{tabular}{ccccccc}
Diagram &Channel            &Mass          &Charge   
   &Term                                      
   &$\beta$       &$\chi$  \\
\hline
c       &$\Sigma^0\, \pi^+$ &$\Sigma \pi$  &$+1$     
   &$f_{\Sigma \Sigma \pi}^2 \, m_{\Sigma \Sigma \pi}$  
   &$-2 \, F^2$
                  &$-2.16$ \\
c       &$\Lambda\, \pi^+$  &$\Lambda\pi$  &$+1$     
   &$f_{\Sigma \Lambda\pi}^2 \, m_{\Sigma \Lambda\pi}$  
   &$-2 \, D^2 / 3$ 
                  &$-1.67$ \\
g,h     &$\Xi^0 \, K^+$     &$\Xi K$       &$+1$     
   &$2 f_{\Sigma \Xi K}^2 \, m_{\Sigma \Xi K}$          
   &$-(F+D)^2$ 
                  &$-6.87$ \\
\end{tabular}
\end{ruledtabular}
%\end{sidewaystable}
%\end{center}

\vspace{18pt}

\caption{Determination of direct $u$-quark sea-quark loop
contributions to the $\Sigma^+$ magnetic moment as illustrated in
Fig.~\protect\ref{SigmaCloud}.}
\label{tab:uSigmaSea}
%\begin{sidewaystable}
%\begin{center}
\begin{ruledtabular}
\begin{tabular}{ccccccc}
Diagram   &Channel         &Mass     &Charge   
   &Term                            
   &$\beta$    &$\chi$  \\
\hline
b       &$\Sigma^0\, \pi^+$ &$\Sigma \pi$  &$-1$     
   &$-f_{\Sigma \Sigma \pi}^2 \, m_{\Sigma \Sigma \pi}$  
   &$2\, F^2$  
                  &$+2.16$ \\
b       &$\Lambda\, \pi^+$  &$\Lambda\pi$  &$-1$     
   &$-f_{\Sigma \Lambda\pi}^2 \, m_{\Sigma \Lambda\pi}$  
   &$2 \, D^2 / 3$  
                  &$+1.67$ \\
e       &$p \overline{K^0}$ &$N K$         &$-1$     
   &$-2f_{\Sigma N K}^2 \, m_{\Sigma N K}$               
   &$(D-F)^2$
                  &$+0.29$ \\
\end{tabular}
\end{ruledtabular}
%\end{sidewaystable}
%\end{center}

\vspace{18pt}

\caption{Indirect sea-quark loop contributions from $u$ valence quarks
to the $\Sigma^+$ magnetic moment.  Here the $u$-valence quark forms a
meson composed with a sea-quark loop as illustrated in
Fig.~\protect\ref{SigmaCloud}.}
\label{tab:uSigmaQuench}
%\begin{sidewaystable}
%\begin{center}
\begin{ruledtabular}
\begin{tabular}{ccccccc}
Diagram   &Channel         &Mass     &Charge   
   &Term                            
   &$\beta$       &$\chi$  \\
\hline 
b       &$\Sigma^0\, \pi^+$ &$\Sigma \pi$  &$+1$     
   &$f_{\Sigma \Sigma \pi}^2 \, m_{\Sigma \Sigma \pi}$  
   &$-2\, F^2$  
                  &$-2.16$ \\
b       &$\Lambda\, \pi^+$  &$\Lambda\pi$  &$+1$     
   &$f_{\Sigma \Lambda\pi}^2 \, m_{\Sigma \Lambda\pi}$  
   &$-2 \, D^2 / 3$  
                  &$-1.67$ \\
c       &$\Sigma^0\, \pi^+$ &$\Sigma \pi$  &$+1$     
   &$f_{\Sigma \Sigma \pi}^2 \, m_{\Sigma \Sigma \pi}$  
   &$-2\, F^2$  
                  &$-2.16$ \\
c       &$\Lambda\, \pi^+$  &$\Lambda\pi$  &$+1$     
   &$f_{\Sigma \Lambda\pi}^2 \, m_{\Sigma \Lambda\pi}$  
   &$-2 \, D^2 / 3$  
                  &$-1.67$ \\
h       &$\Sigma^0\, \pi^+$ &$\Xi K$       &$+1$     
   &$f_{\Sigma \Sigma \pi}^2 \, m_{\Sigma \Xi K}$       
   &$-2\, F^2$  
                  &$-2.16$ \\
h       &$\Lambda\, \pi^+$  &$\Xi K$       &$+1$     
   &$f_{\Sigma \Lambda\pi}^2 \, m_{\Sigma \Xi K}$       
   &$-2 \, D^2 / 3$  
                  &$-1.67$ \\
\end{tabular}
\end{ruledtabular}
%\end{sidewaystable}
%\end{center}
\end{table*}

Tables \ref{tab:uSigmaTotal} through \ref{tab:uSigmaQuench} describe
the various LNA contributions of the $u$ quark to the $\Sigma^+$
magnetic moment derived from Fig.~\ref{SigmaCloud}.  The total
contribution of the $u$ quark alone is isolated by setting the charge
of the $s$ and $d$ quarks to zero, and otherwise using standard
techniques.  The LNA behavior of the $u$-quark contribution to the
$\Sigma^+$ magnetic moment is $-2.16\, m_{\Sigma \Sigma \pi} - 1.67\,
m_{\Sigma \Lambda\pi} -6.87\, m_{\Sigma \Xi K}$.  Sea-quark-loop
contributions are isolated by using meson-baryon couplings where quark
loops of $u$ and $s$ quark flavors (the valence flavors of $\Sigma^+$)
are replaced by a $d$ quark.  Table~\ref{tab:uSigmaSea} summarizes the
direct sea-quark loop contributions.  Subtracting these contributions
leaves a net valence contribution of $-0.29\, m_{\Sigma N K} - 4.32\,
m_{\Sigma \Sigma \pi} - 3.33\, m_{\Sigma \Lambda\pi} -6.87\, m_{\Sigma
\Xi K}$ in full QCD.  Table~\ref{tab:uSigmaQuench} describes indirect
sea-quark loop contributions from $u$ valence quarks in mesons formed
with a sea-quark loop.  Removing these contributions provides the net
quenched $u$-valence contribution of $-0.29\, m_{\Sigma N K} - 3.04\,
m_{\Sigma \Xi K}$.

\begin{table*}[t]
\caption{Determination of the total $d$-quark contribution to the $\Sigma^+$
magnetic moment as illustrated in Fig.~\protect\ref{SigmaCloud}.
$d$-quark contributions are purely sea in origin such that the valence
contribution vanishes.}
\label{tab:dSigmaTotal}
%\begin{sidewaystable}
%\begin{center}
\begin{ruledtabular}
\begin{tabular}{ccccccc}
Diagram &Channel            &Mass          &Charge   
   &Term                                      
   &$\beta$       &$\chi$  \\
\hline
c       &$\Sigma^0\, \pi^+$ &$\Sigma \pi$  &$-1$     
   &$-f_{\Sigma \Sigma \pi}^2 \, m_{\Sigma \Sigma \pi}$  
   &$2\, F^2$  
                  &$+2.16$ \\
c       &$\Lambda\, \pi^+$  &$\Lambda\pi$  &$-1$     
   &$-f_{\Sigma \Lambda\pi}^2 \, m_{\Sigma \Lambda\pi}$  
   &$2 \, D^2 / 3$  
                  &$+1.67$ \\
f       &$p \overline{K^0}$ &$N K$         &$-1$     
   &$-2f_{\Sigma N K}^2 \, m_{\Sigma N K}$               
   &$(D-F)^2$
                  &$+0.29$ \\
\end{tabular}
\end{ruledtabular}
%\end{sidewaystable}
%\end{center}
\end{table*}

The $d$-quark contributions to the LNA behavior of the $\Sigma^+$
magnetic moment are pure sea in origin.  Therefore the total
contributions are the sea contributions such that the valence
$d$-quark contributions vanish.  Table~\ref{tab:dSigmaTotal}
summarizes the contributions.

\begin{table*}[t]
\caption{Determination of the total $s$-quark contribution to the $\Sigma^+$
magnetic moment as illustrated in Fig.~\protect\ref{SigmaCloud}.}
\label{tab:sSigmaTotal}
%\begin{sidewaystable}
%\begin{center}
\begin{ruledtabular}
\begin{tabular}{ccccccc}
Diagram &Channel            &Mass          &Charge   
   &Term                                      
   &$\beta$       &$\chi$  \\
\hline
f       &$p \overline{K^0}$ &$N K$         &$+1$     
   &$2f_{\Sigma N K}^2 \, m_{\Sigma N K}$               
   &$-(D-F)^2$
                  &$-0.29$ \\
g,h     &$\Xi^0 \, K^+$     &$\Xi K$       &$-1$     
   &$-2 f_{\Sigma \Xi K}^2 \, m_{\Sigma \Xi K}$         
   &$(F+D)^2$ 
                  &$+6.87$ \\
\end{tabular}
\end{ruledtabular}
%\end{sidewaystable}
%\end{center}

\vspace{18pt}

\caption{Determination of direct $s$-quark sea-quark loop
contributions to the $\Sigma^+$ magnetic moment as illustrated in
Fig.~\protect\ref{SigmaCloud}.  $\eta_s$ denotes the $s \overline s$
$\eta$ meson.}
\label{tab:sSigmaSea}
%\begin{sidewaystable}
%\begin{center}
\begin{ruledtabular}
\begin{tabular}{ccccccc}
Diagram   &Channel         &Mass     &Charge   
   &Term                            
   &$\beta$    &$\chi$  \\
\hline
h       &$\Sigma^0\, \pi^+$ &$\Xi K$       &$-1$     
   &$-f_{\Sigma \Sigma \pi}^2 \, m_{\Sigma \Xi K}$       
   &$2\, F^2$  
                  &$+2.16$ \\
h       &$\Lambda\, \pi^+$  &$\Xi K$       &$-1$     
   &$-f_{\Sigma \Lambda\pi}^2 \, m_{\Sigma \Xi K}$       
   &$2 \, D^2 / 3$  
                  &$+1.67$ \\
i       &$p \overline{K^0}$ &$\Sigma \eta_s$         &$-1$     
   &$-2f_{\Sigma N K}^2 \, m_{\Sigma \Sigma \eta_s}$               
   &$(D-F)^2$
                  &$+0.29$ \\
\end{tabular}
\end{ruledtabular}
%\end{sidewaystable}
%\end{center}

\vspace{18pt}

\caption{Indirect sea-quark loop contributions from $s$ valence quarks
to the $\Sigma^+$ magnetic moment.  Here the $s$-valence quark forms a
meson composed with a sea-quark loop as illustrated in
Fig.~\protect\ref{SigmaCloud}.  $\eta_s$ denotes the $s \overline s$
$\eta$ meson.}
\label{tab:sSigmaQuench}
%\begin{sidewaystable}
%\begin{center}
\begin{ruledtabular}
\begin{tabular}{ccccccc}
Diagram   &Channel         &Mass     &Charge   
   &Term                            
   &$\beta$       &$\chi$  \\
\hline 
e       &$p \overline{K^0}$ &$N K$         &$+1$     
   &$2f_{\Sigma N K}^2 \, m_{\Sigma N K}$               
   &$-(D-F)^2$
                  &$-0.29$ \\
f       &$p \overline{K^0}$ &$N K$         &$+1$     
   &$2f_{\Sigma N K}^2 \, m_{\Sigma N K}$               
   &$-(D-F)^2$
                  &$-0.29$ \\
i       &$p \overline{K^0}$ &$\Sigma \eta_s$         &$+1$     
   &$2f_{\Sigma N K}^2 \, m_{\Sigma \Sigma \eta_s}$               
   &$-(D-F)^2$
                  &$-0.29$ \\
\end{tabular}
\end{ruledtabular}
%\end{sidewaystable}
%\end{center}
\end{table*}

$s$-quark contributions to the $\Sigma^+$ magnetic moment are
summarized in Tables \ref{tab:sSigmaTotal} through
\ref{tab:sSigmaQuench}.  Removal of the direct sea-quark-loop
contributions from the total contributions provides an $s$-valence
contribution of $-0.29\, m_{\Sigma N K} + 3.04\, m_{\Sigma \Xi K}
-0.29\, m_{\Sigma \Sigma \eta_s}$ in full QCD.  Further removal of the
indirect sea-quark loop contributions of Table \ref{tab:sSigmaQuench}
provides the net quenched $s$-valence contribution of $+0.29\,
m_{\Sigma N K} + 3.04\, m_{\Sigma \Xi K}$.

Charge symmetry provides the quark sector contributions to the
$\Sigma^-$ baryon, while the $\Sigma^0$-baryon results are obtained
from the isospin average of $\Sigma^+$ and $\Sigma^-$.

The $SU(3)$-flavor symmetry of the direct sea-quark-loop contributions
to the $\Sigma^+$ baryon magnetic moment is manifest throughout Tables
\ref{tab:uSigmaSea}, \ref{tab:dSigmaTotal} and \ref{tab:sSigmaSea}.
However, the implementation of $SU(3)$-flavor breaking via the hadron
masses hides the flavor symmetry in the results summarized in
Sec.~\ref{sec:results}, where both meson and baryon mass splittings
are maintained.  $SU(3)$-flavor breaking gives rise to very
different behaviors for these contributions.  This is particularly
true in the common application of holding the strange-quark mass fixed
while varying the light $u$ and $d$ masses.  In this case the
$\eta_{s}$ meson mass is constant.

\end{subsection}

\begin{subsection}{$\Lambda$ and $\Xi$ Baryons }
\label{subsec:lambdaXi}

The derivation of the quark sector contributions to $\Xi$ baryons
proceeds in precisely the same manner as that for the $\Sigma$
baryons.  As there are no new concepts, derivation is left as an
exercise for the interested reader.  $\Xi$-baryon results are
summarized in Sec.~\ref{sec:results}.

However, the flavor singlet structure of the $\Lambda$ baryon presents
a problem to the approach described thus far.  The necessary presence
of $u$-, $d$- and $s$-quark flavors simultaneously, appears to require
the introduction of a fourth quark flavor and its associated $SU(4)$
couplings to describe the disconnected sea-quark loop contributions.

Fortunately one can exploit the $SU(3)$-flavor symmetry relation
among octet baryons.  Such a relation is manifest in two- and
three-point correlation functions for the $\Lambda$
\cite{Leinweber:1990dv}.  Denoting the two-point correlation function
for $\Sigma^0$ as $\Sigma^0_s(x)$, one has
\begin{equation}
\Lambda(x) = {1 \over 3} \left ( 2\, \Sigma^0_u(x) + 2\, \Sigma^0_d(x)
- \Sigma^0_s(x) \right ) \, ,
\end{equation} 
where $\Sigma^0_s(x)$ has symmetry between $u$ and $d$ quarks,
$\Sigma^0_u(x)$ has symmetry between $s$ and $d$ quarks, and similarly
$\Sigma^0_d(x)$ has symmetry between $u$ and $s$ quarks.  Just as
\begin{equation}
\Sigma^0_s(x) = {1 \over 2} \left ( \Sigma^+(x) + \Sigma^-(x) \right ) \, ,
\end{equation}
one also has
\begin{equation}
\Sigma^0_u(x) = {1 \over 2} \left ( n(x) + \Xi^0(x) \right ) \, ,
\end{equation}
and
\begin{equation}
\Sigma^0_d(x) = {1 \over 2} \left ( p(x) + \Xi^-(x) \right ) \, ,
\end{equation}
in the $SU(3)$-flavor limit, such that
\begin{eqnarray}
\Lambda(x) &=& {1 \over 3} \biggl ( 
p(x) + n(x) + \Xi^0(x) + \Xi^-(x) \nonumber \\
&& \qquad -{1\over 2} \left [ \Sigma^+(x) + \Sigma^-(x) \right ] 
\biggr ) \, .
\label{SU3lambda}
\end{eqnarray} 
Note that this relation holds for any electric-charge assignments to
the quark flavors, such that individual quark-flavor contributions
can be resolved.

Of course it is essential to recover $SU(3)$-flavor violations.  To
do this one begins exactly as for the proton or $\Sigma^+$ described
above by constructing the quark-flow diagrams describing the one-loop
meson cloud of the $\Lambda$.  The couplings of all sea-quark loop
contributions can be related to the three quark-flow diagrams of
Fig.~\ref{LambdaCloud}.  Unknown couplings $f_u^2$, $f_d^2$ and
$f_s^2$ are introduced to describe the couplings of diagrams (a), (b)
and (c) respectively.  Our working approximation of exact
$SU(2)$-isospin symmetry at the current-quark level provides $f_d^2 =
f_u^2$ in $\Lambda$, leaving two parameters, $f_u^2$ and $f_s^2$, to be
determined via the $SU(3)$ relation of Eq.~(\ref{SU3lambda}).  As both
the light- and strange-quark contributions to the $\Lambda$ moment can
be resolved, there are two $SU(3)$ relations to constrain the two
parameters $f_u^2$ and $f_s^2$.  This is particularly easy, when one
recalls that the indirect sea-quark loop contribution from a $u$ or
$s$ valence quark participating in a meson constructed with a
sea-quark loop are proportional to either $f_u^2$ or $f_s^2$ alone.
Results are summarized in Sec.~\ref{sec:results}.

\begin{figure}[b]
\centering{\
\epsfig{file=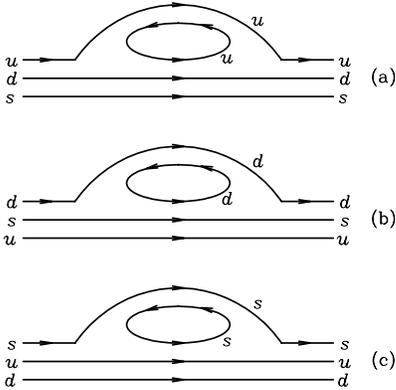,width=0.6\hsize,angle=0} }
\caption{Key quark-flow diagrams for the $\Lambda$ baryon.  Diagrams (a),
  (b) and (c) are proportional to the introduced couplings $f_u^2$,
  $f_d^2$ and $f_s^2$ respectively.  }
\label{LambdaCloud}
\end{figure}

\end{subsection}

%
%======== ========= ========= ========= ========= ========= ========= =========
%

\begin{subsection}{Quenched Exotics}
\label{subsec:eta}

\begin{figure}[t]
\centering{\
\epsfig{file=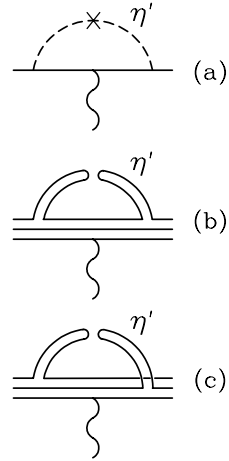,width=0.7\hsize,angle=90} }
\caption{Diagrams giving rise to the logarithmic divergence of a baryon
  magnetic moment in the quenched approximation.  The cross on the
  meson propagator in (a) denotes the double hairpin graph of the
  quark-flow diagrams of (b) and (c).}
\label{EtaPrimeLog}
\end{figure}

The double hair-pin graph of Fig.~\ref{EtaPrimeLog} associated with
the quenched-$\eta'$ meson gives rise to new singular $\log(m_\pi)$
behavior in the chiral limit \cite{Savage:2001dy}.  This logarithmic
term provides a correction to the tree-level term.  The contribution
has its origin in the loop integral of Fig.~\ref{EtaPrimeLog}(a)
corresponding to
\begin{equation}
-i {16\pi^2\over 3}
\int {d^4 q\over (2\pi)^4}
{q^2-(v\cdot q)^2\over
[v\cdot q + i \epsilon]^2
[q^2-m_1^2 + i \epsilon][q^2-m_2^2 + i \epsilon]} \, .
\end{equation}
For equal singlet-meson masses $m_1 = m_2$, as in the quark flow of
Fig.~\ref{EtaPrimeLog}(b), this integral provides the nonanalytic
behavior of 
\begin{equation}
\log\left( {m^2\over\Lambda^2}\right)
\, .
\end{equation}
However, Fig.~\ref{EtaPrimeLog}(c) indicates that the meson masses in
the double hair-pin graph need not be equal when considering hyperon
magnetic moments.  Here the $\eta'$-meson masses $m_1$ and $m_2$ may
correspond to different quark-antiquark sources; e.g.\ $\eta'(u
\overline u)$ versus $\eta'(s \overline s)$.  When $m_1 \ne m_2$, one
finds a nonanalytic contribution of
\begin{equation}
{m_1^2\log \left({m_1^2\over\Lambda^2}\right)
-m_2^2\log \left({m_2^2\over\Lambda^2}\right)\over m_1^2-m_2^2}
\, .
\end{equation}

Hence, for the hyperons, one must isolate the doubly- and
singly-represented quark sector couplings to $\eta'$ mesons.  Consider
for example, $\eta'$ couplings for $\Sigma^+$ involving $u$ quarks.
The transition $\Sigma^+ \to \Sigma^+ \pi^0$ involves $u$ quarks alone
at one loop and can be used to determine the $\eta^\prime_u$ coupling.
The $\Sigma^+$-$\pi^0$ coupling is $f_{\Sigma \Sigma \pi} = 2\, F$.
Since
\begin{eqnarray*}
\mid\! \pi^0 \,\rangle &=& {1 \over \sqrt 2} 
\left ( \mid\! u \overline u \,\rangle - \mid\! d \overline d \,\rangle \right ) \, , \\
\mid\! \eta \,\rangle &=& {1 \over \sqrt 6} 
\left ( \mid\! u \overline u \,\rangle + \mid\! d \overline d \,\rangle - 2\, \mid\! s \overline s \,\rangle \right ) \, ,
\\
\mid\! \eta' \,\rangle &=& {1 \over \sqrt 3} 
\left ( \mid\! u \overline u \,\rangle + \mid\! d \overline d \,\rangle + \mid\! s \overline s \,\rangle \right ) \, ,
\\
\end{eqnarray*}
one has
\begin{equation}
\mid\! u \overline u \,\rangle = {1 \over \sqrt 6} 
\left ( \sqrt 3\,  \mid\! \pi^0 \,\rangle + \mid\! \eta \,\rangle + \sqrt 2\, \mid\! \eta' \,\rangle \right ) \, .
\end{equation}
The $\eta^\prime_u$ coupling is $\sqrt{2/3}$ of the pion coupling to
$\mid\! u \overline u \,\rangle$; i.e. $2 \sqrt{2/3} \, F$.  

For the singly represented quark sector, consider $\Xi^0 \to \Xi^0
\pi^0$ involving $u$-quarks alone at one loop.  Here the $u$-quark
coupling is $f_{\Xi \Xi \pi} = (F - D)$, such that the $\eta^\prime_u$
coupling to $\Xi$ baryons is $\sqrt{2/3}\, (F - D)$.  

To check this separation of quark sector contributions to
$\eta^\prime$ contributions, consider the proton.  Here the
contribution to intermediate $\eta'$ states is
\begin{eqnarray}
&& {2 \over 3} \left \{
4\, F^2 \, I(\eta_{u},\eta_{u}) + 
2 \cdot 2 F\, (F - D) I(\eta_{u},\eta_{d})
\right . \nonumber \\
&& \left . \quad\:
 + (F - D)^2 I(\eta_{d},\eta_{d}) \right \} \, ,
\label{TwoFlavors}
\end{eqnarray}%
where $I(\eta_{u},\eta_{d})$ denotes the
loop integral of Fig.~\ref{EtaPrimeLog}(a) for quark flow diagram
Fig.~\ref{EtaPrimeLog}(c).  For equal $\eta_{u}$ and
$\eta_{d}$ masses one recovers
\begin{equation}
{2 \over 3} (3F - D)^2 \, \log\left( {m^2\over\Lambda^2}\right) \, ,
\end{equation}
where the leading factor is the standard $NN\eta'$ coupling.
Double-hairpin $\eta'$ contributions to $\Sigma$ and $\Xi$ baryons may
be obtained from Eq.~(\ref{TwoFlavors}) with the appropriate
quark-flavor assignments.  For example, the factor multiplying the
tree level contribution to $\Xi^-$-baryon quark-sector magnetic
moments is
\begin{eqnarray}
1 &-& \xi_0 \, {2 \over 3} \left \{
4\, F^2 \, \log\left( {m_{s \overline s}^2 \over \Lambda^2}\right)
\right . 
\nonumber \\
&& \quad \: +
2 \cdot 2 F\, (F - D) 
{m_{s \overline s}^2\log \left({m_{s \overline s}^2\over\Lambda^2}\right)
-m_\pi^2 \log \left({m_\pi^2 \over \Lambda^2} \right) \over m_{s
\overline s}^2-m_\pi^2}
\nonumber \\
&& \left . \quad\:
 + (F - D)^2 \, \log\left( {m_\pi^2 \over  \Lambda^2}\right)
\right \} \, ,
\label{XiTwoFlavors}
\end{eqnarray}
where remaining loop-integral factors have been incorporated in
\begin{equation}
\xi_0 = { M_0^2 \over 16 \, \pi^2 \, f_\pi^2 }\, ,
\end{equation}
with the double hair-pin interaction strength $M_0 \sim 0.75$ GeV
\cite{Golterman:1994,Kuramashi:1994aj}.  While this logarithmic
divergence dominates the chiral expansion near the chiral limit,
application of these results to the extrapolation of the quenched
proton magnetic moment \cite{Young:2003} reveals that the curvature
associated with this term is small for $m_\pi^2 \agt 0.1\
\mbox{GeV}^2$.

\end{subsection}

\end{section}

%%%%%%%%%%%%%%%%%%%%%%%%%%%%%%%%%%%%%%%%%%%%%%%%%%%%%%%%%%%%%%%%%%%%%%%%%%%

\begin{section}{RESULTS}
\label{sec:results}

This approach allows one to separate an individual quark-flavor
contribution to a baryon form factor into five categories, namely:
``total'' full-QCD contributions, ``direct sea-quark loop,'' and
``valence'' contributions of full-QCD, obtained by removing the direct
current coupling to sea-quark loops from the total contributions.
Upon further removing ``indirect sea-quark loop'' contributions, one
obtains the ``quenched valence''
contributions. Tables~\ref{tab:quarkMomBetaA} and
\ref{tab:quarkMomBetaB} report the axial couplings for these
quark-sector contributions to baryon magnetic moments.

The LNA ``direct sea-quark loop'' contribution is relevant
to disconnected insertions of the EM current in {\it either} full or
quenched QCD, whereas the LNA ``valence'' contribution is relevant to
connected insertions of the EM current only in full QCD.  The final
category of ``quenched valence'' contributions is relevant to connected
insertions of the EM current in quenched QCD.  The latter is commonly
referred to as the quenched QCD result.  

The channels denoted $\Omega K$ in Tables~\ref{tab:quarkMomBetaA},
\ref{tab:quarkMomBetaB} and in the following actually involve the
propagation of an octet $sss$ baryon; i.e.\ the $\Xi^-$ baryon with
$m_d = m_s$.  In separating valence and sea-quark loop contributions,
the cancellation of valence and sea-quark loop octet-$sss$-baryon
contributions does not occur.  Figure~\ref{fig:sss} provides quark
flow diagrams for $\Xi^0 \to \Omega^- K^+$ which illustrate this
phenomenon.

\begin{figure}[t]
\centering{\
\epsfig{file=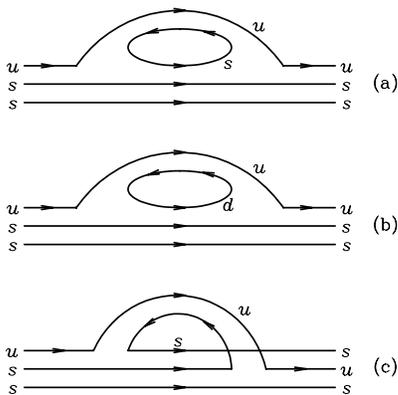,width=0.6\hsize,angle=0} }
\caption{Quark-flow diagrams illustrating the presence of an
  $sss$-octet baryon propagating in each of diagrams (a) and (c).
  Further discussion is provided in the text.}
\label{fig:sss}
\end{figure}

In the $SU(3)$-flavor symmetry limit, Figs.~\ref{fig:sss}(a) and (b)
are equivalent due to the flavor-blindness of QCD interactions.
Fig.~\ref{fig:sss}(b) certainly has overlap with an octet-$\Xi^-\,
\pi^+$ intermediate state.  Hence the diagram of Fig.~\ref{fig:sss}(a)
also has an octet baryon propagating in the intermediate state.  Of
course, we know there is no $sss$ octet baryon and this problem is
solved by the contribution of Fig.~\ref{fig:sss}(c) which must be
equal but opposite in sign to Fig.~\ref{fig:sss}(a) when an octet
baryon propagates in the intermediate state, thus eliminating the
octet $sss$ baryon contribution in full QCD.  In separating valence
and sea contributions, each quark flow graph must be taken on its own
such that octet baryons are not necessarily eliminated.  Indeed, in
quenched QCD, only Fig.~\ref{fig:sss}(c) survives, and this quark flow
graph has an $sss$-octet baryon propagating in the intermediate state.
To some extent this physics has already been seen in
Figs.~\ref{ProtonCloud}(d) plus (e) where only a decuplet baryon can
contribute in full QCD, but octet baryons provide contributions in the
process of separating valence and sea sectors.

Baryon moments are constructed from the quark sector coefficients by
multiplying the $u$, $d$ and $s$ results by their appropriate charge
factors and summing.  For example, the proton moment is 
\begin{equation}
\mu_p = {2 \over 3} u_p  - {1 \over 3} d_p - {1 \over 3} s_p \, ,
\end{equation}
and the neutron moment is 
\begin{equation}
\mu_n = -{1 \over 3} u_p + {2 \over 3} d_p - {1 \over 3} s_p \, .
\end{equation}
Similarly, the $\Sigma^+$ moment is 
\begin{equation}
\mu_{\Sigma^+} = {2 \over 3} u_{\Sigma^+}  - {1 \over 3} d_{\Sigma^+}
- {1 \over 3} s_{\Sigma^+} \, , 
\end{equation}
and the $\Sigma^-$ moment is 
\begin{equation}
\mu_{\Sigma^-} = -{1 \over 3} u_{\Sigma^+} + {2 \over 3} d_{\Sigma^+}
- {1 \over 3} s_{\Sigma^+} \, . 
\end{equation}
Tables~\ref{tab:baryonMomBetaA} and \ref{tab:baryonMomBetaB} report
the axial couplings for the intermediate meson-baryon channels
contributing to the nonanalytic behavior of baryon magnetic moments.
We note that upon neglecting the baryon mass splittings, one recovers
the full QCD results of Ref.~\cite{Jenkins:1992pi} summarized in their
Eqs.~(A.2) and (A.4), and the quenched results of
Ref.~\cite{Savage:2001dy} summarized in their Table 2.

Table~\ref{tab:quarkMomChi} reports values for the coefficient,
$\chi$, providing the LNA contribution to baryon magnetic moments
($\chi \, m_\pi$ or $\chi \, m_K$ or $\chi \, m_{\eta_s}$ as
appropriate) by quark sectors with each quark flavor normalized to
unit charge.  Charge symmetry provides the contributions for other
baryons.  Values are based on the tree-level axial couplings $F =
0.50$ and $D = 0.76$ with $f_\pi = 93$ MeV.  Similar results for bulk
baryon moments are provided in Table~\ref{tab:baryonMomChi}.  For
convenience, values using the one-loop corrected values
\cite{Jenkins:1992pi} of $F = 0.40$ and $D = 0.61$ are provided in
Tables~\ref{tab:quarkMomChiOneLoop} and \ref{tab:baryonMomChiOneLoop}
respectively.

\begin{turnpage}
%\begin{longtable*}[t]
\begin{table*}[p]
%\squeezetable
\caption{Coefficients, $\beta$, providing the LNA contribution to
nucleon, $\Sigma$ and $\Lambda$-baryon magnetic moments by quark
sectors with quark charges normalized to unit charge.  Intermediate
(Int.) meson-baryon channels are indicated to allow for
$SU(3)$-flavor breaking in both the meson and baryon masses.  Total,
direct sea-quark loop, valence, indirect sea-quark loop and quenched valence
coefficients are indicated. }
\label{tab:quarkMomBetaA}
\addtolength{\tabcolsep}{-1pt}
%\begin{tabular}{lllllll}
\begin{ruledtabular}
%\hline
%\hline
\begin{tabular}{ccccccc}
$q$   &Int.           &Total Quark Sector    
                      &Direct Sea-Quark Loop     
                      &Valence Sector     
                      &Indirect Loop
                      &Quenched Valence \\
\noalign{\smallskip}
\hline 
\noalign{\smallskip}
$u_p$ &$N \pi$        &$- (D + F)^2$
                      &$(5\, D^2 - 6\, D\, F + 9\, F^2)/3$
                      &$- 4\, (2\, D^2 + 3\, F^2)/3$
                      &$- 4\, (D^2 + 3\, F^2)/3$
                      &$- 4\, D^2/3$ \\     
      &$\Lambda K$    &$- (D + 3\, F)^2/6$
                      &$0$
                      &$- (D + 3\, F)^2/6$
                      &$- (D + 3\, F)^2/6$
                      &$0$ \\     
      &$\Sigma  K$    &$- (D - F)^2/2 $
                      &$0$
                      &$- (D - F)^2/2$
                      &$- (D - F)^2/2$
                      &$0$ \\     
\noalign{\smallskip}   
$d_p$ &$N \pi$        &$(D + F)^2$
                      &$(5\, D^2 - 6\, D\, F + 9\, F^2)/3$
                      &$- 2\, (D^2 - 6\, D\, F + 3\, F^2)/3$
                      &$- 2\, (D - F)^2$
                      &$4\, D^2/3$ \\     
      &$\Sigma  K$    &$- (D - F)^2$
                      &$0$
                      &$- (D - F)^2$
                      &$- (D - F)^2$
                      &$0$ \\     
\noalign{\smallskip}   
$s_p$ &$\Lambda K$    &$(D + 3\, F)^2/6$
                      &$(D + 3\, F)^2/6$
                      &$0$
                      &$0$
                      &$0$ \\     
      &$\Sigma K$     &$3\, (D - F)^2/2$
                      &$3\, (D - F)^2/2$
                      &$0$
                      &$0$
                      &$0$ \\     
\noalign{\smallskip}
\hline 
\noalign{\smallskip}
$u_{\Sigma^+}$   &$\Sigma  \pi$  &$- 2\, F^2$
                                 &$  2\, F^2$
                                 &$- 4\, F^2$
                                 &$- 4\, F^2$
                                 &$0$ \\     
                 &$\Lambda \pi$  &$- 2\, D^2 / 3$
                                 &$  2\, D^2 / 3$
                                 &$- 4\, D^2 / 3$
                                 &$- 4\, D^2 / 3$
                                 &$0$ \\     
                 &$N K        $  &$0$
                                 &$(D - F)^2$
                                 &$- (D - F)^2$
                                 &$0$
                                 &$- (D - F)^2$ \\     
                 &$\Xi K      $  &$- (D + F)^2$
                                 &$0$
                                 &$- (D + F)^2$
                                 &$- 2\, (D^2 + 3\, F^2)/3$
                                 &$-(D^2 + 6\, D\, F - 3\, F^2)/3$ \\     
\noalign{\smallskip}		  
$d_{\Sigma^+}$   &$\Sigma  \pi$  &$2\, F^2$
                                 &$2\, F^2$
                                 &$0$
                                 &$0$
                                 &$0$ \\     
                 &$\Lambda \pi$  &$2\, D^2 / 3$
                                 &$2\, D^2 / 3$
                                 &$0$
                                 &$0$
                                 &$0$ \\     
                 &$N K        $  &$(D - F)^2$
                                 &$(D - F)^2$
                                 &$0$
                                 &$0$
                                 &$0$ \\     
\noalign{\smallskip}		  
$s_{\Sigma^+}$   &$\Sigma \eta_{s}$  
                                 &$0$
                                 &$(D - F)^2$
                                 &$- (D - F)^2$
                                 &$- (D - F)^2$
                                 &$0$ \\     
                 &$N K        $  &$- (D - F)^2$
                                 &$0$
                                 &$- (D - F)^2$
                                 &$- 2\, (D - F)^2$
                                 &$(D - F)^2$ \\     
                 &$\Xi K      $  &$(D + F)^2$
                                 &$2\, (D^2 + 3\, F^2) / 3$
                                 &$(D^2 + 6\, D\, F - 3\, F^2) / 3$
                                 &$0$
                                 &$(D^2 + 6\, D\, F - 3\, F^2) / 3$ \\     
\noalign{\smallskip}		  
$u_{\Sigma^0} \mid d_{\Sigma^0}$   
                 &$\Sigma  \pi$  &$0$
                                 &$  2\, F^2$
                                 &$- 2\, F^2$
                                 &$- 2\, F^2$
                                 &$0$ \\     
                 &$\Lambda \pi$  &$0$
                                 &$  2\, D^2 / 3$
                                 &$- 2\, D^2 / 3$
                                 &$- 2\, D^2 / 3$
                                 &$0$ \\     
                 &$N K        $  &$(D - F)^2 / 2$
                                 &$(D - F)^2$
                                 &$- (D - F)^2 / 2$
                                 &$0$
                                 &$- (D - F)^2 / 2$ \\     
                 &$\Xi K      $  &$- (D + F)^2 / 2$
                                 &$0$
                                 &$- (D + F)^2 / 2$
                                 &$- (D^2 + 3\, F^2) / 3$
                                 &$-(D^2 + 6\, D\, F - 3\, F^2) / 6$ \\     
\noalign{\smallskip}
\hline 
\noalign{\smallskip}
$u_\Lambda \mid d_\Lambda$
              &$\Sigma \pi $     &$0$
                                 &$  2\, D^2 / 3$
                                 &$- 2\, D^2 / 3$
                                 &$- 2\, D^2 / 3$
                                 &$0$ \\     
              &$\Lambda \eta_{l}$  
                                 &$0$
                                 &$2\, (2\, D - 3\, F)^2 / 9$
                                 &$- 2\, (2\, D - 3\, F)^2 / 9$
                                 &$- 2\, (2\, D - 3\, F)^2 / 9$
                                 &$0$ \\     
              &$N K        $     &$(D + 3\, F)^2 / 6$
                                 &$(D + 3\, F)^2 / 9$
                                 &$(D + 3\, F)^2 / 18$
                                 &$0$
                                 &$(D + 3\, F)^2 / 18$ \\     
              &$\Xi K      $     &$- (D - 3\, F)^2 / 6$
                                 &$0$
                                 &$- (D - 3\, F)^2 / 6$
                                 &$( - 7\, D^2 + 12\, D\, F - 9\, F^2) / 9$
                                 &$(11\, D^2 - 6\, D\, F - 9\, F^2) / 18$ \\     
\noalign{\smallskip}		  
$s_\Lambda$   &$\Lambda\eta_{s}$   
                                 &$0$
                                 &$(D + 3\, F)^2 / 9$
                                 &$- (D + 3\, F)^2 / 9$
                                 &$- (D + 3\, F)^2 / 9$
                                 &$0$ \\     
              &$N K        $     &$- (D + 3\, F)^2 / 3$
                                 &$0$
                                 &$- (D + 3\, F)^2 / 3$
                                 &$- 2\, (D + 3\, F)^2 / 9$
                                 &$- (D + 3\, F)^2 / 9$ \\     
              &$\Xi K      $     &$(D - 3\, F)^2 / 3$
                                 &$2\, (7\, D^2 - 12\, D\, F + 9\, F^2) / 9$
                                 &$-(11\, D^2 - 6\, D\, F - 9\, F^2) / 9$
                                 &$0$
                                 &$-(11\, D^2 - 6\, D\, F - 9\, F^2) / 9$ \\     
\end{tabular}
\end{ruledtabular}
\end{table*}
%\hline
%\hline
%\end{longtable*}
\end{turnpage}

\begin{turnpage}
\begingroup
%\begin{longtable*}[t]
\begin{table*}[p]
%\squeezetable
\caption{Coefficients, $\beta$, providing the LNA contribution to
$\Xi$-baryon magnetic moments by quark sectors with quark charges
normalized to unit charge.  Intermediate (Int.) meson-baryon channels
are indicated to allow for $SU(3)$-flavor breaking in both the meson
and baryon masses.  Total, direct sea-quark loop, valence, indirect
sea-quark loop and quenched valence coefficients are indicated. }
\label{tab:quarkMomBetaB}
\addtolength{\tabcolsep}{-1pt}
%\begin{tabular}{lllllll}
\begin{ruledtabular}
%\hline
%\hline
\begin{tabular}{ccccccc}
$q$   &Int.           &Total Quark Sector    
                      &Direct Sea-Quark Loop     
                      &Valence Sector     
                      &Indirect Loop
                      &Quenched Valence \\
\noalign{\smallskip}
\hline 
\noalign{\smallskip}
$u_{\Xi^0}$      &$\Xi \pi$      &$- (D - F)^2$
                                 &$(D - F)^2$
                                 &$- 2\, (D - F)^2$
                                 &$- 2\, (D - F)^2$
                                 &$0$ \\     
                 &$\Lambda K$    &$0$
                                 &$(D - 3\, F)^2 / 6$
                                 &$- (D - 3\, F)^2 / 6$
                                 &$0$
                                 &$- (D - 3\, F)^2 / 6$ \\     
                 &$\Sigma K$     &$(D + F)^2$
                                 &$(D + F)^2 / 2$
                                 &$(D + F)^2 / 2$
                                 &$0$
                                 &$(D + F)^2 / 2$ \\     
                 &$\Omega K$     &$0$
                                 &$0$
                                 &$0$
                                 &$- (D - F)^2$
                                 &$(D - F)^2$ \\     
\noalign{\smallskip}	       
$d_{\Xi^0}$      &$\Xi \pi$      &$(D - F)^2$
                                 &$(D - F)^2$
                                 &$0$
                                 &$0$
                                 &$0$ \\     
                 &$\Lambda K$    &$(D - 3\, F)^2 / 6$
                                 &$(D - 3\, F)^2 / 6$
                                 &$0$
                                 &$0$
                                 &$0$ \\     
                 &$\Sigma K$     &$(D + F)^2 / 2$
                                 &$(D + F)^2 / 2$
                                 &$0$
                                 &$0$
                                 &$0$ \\     
\noalign{\smallskip}	       
$s_{\Xi^0}$      &$\Lambda K$    &$- (D - 3\, F)^2 / 6$
                                 &$0$
                                 &$- (D - 3\, F)^2 / 6$
                                 &$- (D - 3\, F)^2 / 3$
                                 &$(D - 3\, F)^2 / 6$ \\     
                 &$\Sigma K$     &$- 3\, (D + F)^2 / 2$
                                 &$0$
                                 &$- 3\, (D + F)^2 / 2$
                                 &$- (D + F)^2$
                                 &$- (D + F)^2 / 2$ \\     
                 &$\Omega K$     &$0$
                                 &$(D - F)^2$
                                 &$- (D - F)^2$
                                 &$0$
                                 &$- (D - F)^2$ \\     
                 &$\Xi \eta_{s}$     
                                 &$0$
                                 &$2\, (D^2 + 3\, F^2) / 3$
                                 &$- 2\, (D^2 + 3\, F^2) / 3$
                                 &$- 2\, (D^2 + 3\, F^2) / 3$
                                 &$0$ \\     
\end{tabular}
\end{ruledtabular}
%\end{table*}
%\hline
%\hline
%\end{longtable*}

\vspace{48pt}

%\begin{longtable*}[t]
%\begin{table*}[p]
%\squeezetable
\caption{Coefficients, $\beta$, providing the LNA contribution to
nucleon magnetic moments.  Intermediate (Int.)  meson-baryon channels
are indicated to allow for $SU(3)$-flavor breaking in both the meson
and baryon masses.  }
\label{tab:baryonMomBetaA}
%\addtolength{\tabcolsep}{-1pt}
\begin{ruledtabular}
%\hline
%\hline
\begin{tabular}{ccccccc}
Baryon   &Int.          &Total Quark Sector    
                        &Direct Sea-Quark Loop     
                        &Valence Sector     
                        &Indirect Loop
                        &Quenched Valence \\
\hline 
\noalign{\smallskip}
$p$   &$N \pi$        &$- (D + F)^2$
                      &$(5\, D^2 - 6\, D\, F + 9\, F^2) / 9$
                      &$2\, ( - 7\, D^2 - 6\, D\, F - 9\, F^2) / 9$
                      &$- 2\, (D + 3\, F)^2 / 9$
                      &$- 4\, D^2/3$ \\     
      &$\Lambda K$    &$- (D + 3\, F)^2 / 6$
                      &$- (D + 3\, F)^2 / 18$
                      &$- (D + 3\, F)^2 / 9$
                      &$- (D + 3\, F)^2 / 9$
                      &$0$ \\     
      &$\Sigma  K$    &$- (D - F)^2 / 2 $
                      &$- (D - F)^2 / 2$
                      &$0$
                      &$0$
                      &$0$ \\     
\noalign{\smallskip}   
$n$   &$N \pi$        &$(D + F)^2$
                      &$(5\, D^2 - 6\, D\, F + 9\, F^2) / 9$
                      &$4\, D\, (D + 6\, F) / 9$
                      &$8\, D\, ( - D + 3\, F) / 9$
                      &$4\, D^2 / 3$ \\     
      &$\Lambda K$    &$0$
                      &$- (D + 3\, F)^2 / 18$
                      &$(D + 3\, F)^2 / 18$
                      &$(D + 3\, F)^2 / 18$
                      &$0$ \\     
      &$\Sigma  K$    &$- (D - F)^2$
                      &$- (D - F)^2 / 2$
                      &$- (D - F)^2 / 2$
                      &$- (D - F)^2 / 2$
                      &$0$ \\     
\end{tabular}
\end{ruledtabular}
\end{table*}
%\hline
%\hline
%\end{longtable*}
\endgroup
\end{turnpage}

\begin{turnpage}
%\begin{longtable*}[t]
\begin{table*}[p]
%\squeezetable
\caption{Coefficients, $\beta$, providing the LNA contribution to
$\Sigma$-, $\Lambda$- and $\Xi$-baryon magnetic moments.  Intermediate
(Int.)  meson-baryon channels are indicated to allow for $SU(3)$-flavor
breaking in both the meson and baryon masses.  }
\label{tab:baryonMomBetaB}
%\addtolength{\tabcolsep}{-1pt}
\begin{ruledtabular}
%\hline
%\hline
\begin{tabular}{ccccccc}
Baryon   &Int.          &Total Quark Sector    
                        &Direct Sea-Quark Loop     
                        &Valence Sector     
                        &Indirect Loop
                        &Quenched Valence \\
\noalign{\smallskip}
\hline 
\noalign{\smallskip}
$\Sigma^+$       &$\Sigma  \pi$  &$- 2\, F^2$
                                 &$  2\, F^2 / 3$
                                 &$- 8\, F^2 / 3$
                                 &$- 8\, F^2 / 3$
                                 &$0$ \\     
                 &$\Lambda \pi$  &$- 2\, D^2 / 3$
                                 &$  2\, D^2 / 9$
                                 &$- 8\, D^2 / 9$
                                 &$- 8\, D^2 / 9$
                                 &$0$ \\     
                 &$N K        $  &$0$
                                 &$(D - F)^2 / 3$
                                 &$- (D - F)^2 / 3$
                                 &$2\, (D - F)^2 / 3$
                                 &$- (D - F)^2$ \\     
                 &$\Xi K      $  &$- (D + F)^2$
                                 &$- 2\, (D^2 + 3\, F^2) / 9$
                                 &$( - 7\, D^2 - 18\, D\, F - 3\, F^2) / 9$
                                 &$- 4\, (D^2 + 3\, F^2) / 9$
                                 &$-(D^2 + 6\, D\, F - 3\, F^2) / 3$ \\     
                 &$\Sigma \eta_{s}$  
                                 &$0$
                                 &$- (D - F)^2 / 3$
                                 &$(D - F)^2 / 3$
                                 &$(D - F)^2 / 3$
                                 &$0$ \\     
\noalign{\smallskip}		  
$\Sigma^0$       &$\Sigma  \pi$  &$0$
                                 &$  2\, F^2 / 3$
                                 &$- 2\, F^2 / 3$
                                 &$- 2\, F^2 / 3$
                                 &$0$ \\     
                 &$\Lambda \pi$  &$0$
                                 &$  2\, D^2 / 9$
                                 &$- 2\, D^2 / 9$
                                 &$- 2\, D^2 / 9$
                                 &$0$ \\     
                 &$N K        $  &$(D - F)^2 / 2$
                                 &$(D - F)^2 / 3$
                                 &$(D - F)^2 / 6$
                                 &$2\, (D - F)^2 / 3$
                                 &$- (D - F)^2 / 2$ \\     
                 &$\Xi K      $  &$- (D + F)^2 / 2$
                                 &$- 2\, (D^2 + 3\, F^2) / 9$
                                 &$( - 5\, D^2 - 18\, D\, F + 3\, F^2) / 18$
                                 &$- (D^2 + 3\, F^2) / 9$
                                 &$-(D^2 + 6\, D\, F - 3\, F^2) / 6$ \\     
                 &$\Sigma \eta_{s}$  
                                 &$0$
                                 &$- (D - F)^2 / 3$
                                 &$(D - F)^2 / 3$
                                 &$(D - F)^2 / 3$
                                 &$0$ \\     
\noalign{\smallskip}		  
$\Sigma^-$       &$\Sigma  \pi$  &$  2\, F^2$
                                 &$  2\, F^2 / 3$
                                 &$  4\, F^2 / 3$
                                 &$  4\, F^2 / 3$
                                 &$0$ \\     
                 &$\Lambda \pi$  &$  2\, D^2 / 3$
                                 &$  2\, D^2 / 9$
                                 &$  4\, D^2 / 9$
                                 &$  4\, D^2 / 9$
                                 &$0$ \\     
                 &$N K        $  &$(D - F)^2$
                                 &$(D - F)^2 / 3$
                                 &$2\, (D - F)^2 / 3$
                                 &$2\, (D - F)^2 / 3$
                                 &$- (D - F)^2$ \\     
                 &$\Xi K      $  &$0$
                                 &$- 2\, (D^2 + 3\, F^2) / 9$
                                 &$2\, (D^2 + 3\, F^2) / 9$
                                 &$2\, (D^2 + 3\, F^2) / 9$
                                 &$0$ \\     
                 &$\Sigma \eta_{s}$  
                                 &$0$
                                 &$- (D - F)^2 / 3$
                                 &$(D - F)^2 / 3$
                                 &$(D - F)^2 / 3$
                                 &$0$ \\     
\noalign{\smallskip}
\hline 
\noalign{\smallskip}
$\Lambda$     &$\Sigma \pi $     &$0$
                                 &$  2\, D^2 / 9$
                                 &$- 2\, D^2 / 9$
                                 &$- 2\, D^2 / 9$
                                 &$0$ \\     
              &$\Lambda \eta_{l}$  
                                 &$0$
                                 &$2\, (2\, D - 3\, F)^2 / 27$
                                 &$- 2\, (2\, D - 3\, F)^2 / 27$
                                 &$- 2\, (2\, D - 3\, F)^2 / 27$
                                 &$0$ \\
              &$N K        $     &$(D + 3\, F)^2 / 6$
                                 &$(D + 3\, F)^2 / 27$
                                 &$7\, (D + 3\, F)^2 / 54$
                                 &$2\, (D + 3\, F)^2 / 27$
                                 &$(D + 3\, F)^2 / 18$ \\     
              &$\Xi K      $     &$- (D - 3\, F)^2 / 6$
                                 &$2\, ( - 7\, D^2 + 12\, D\, F - 9\, F^2) / 27$
                                 &$(19\, D^2 + 6\, D\, F - 45\, F^2) / 54$
                                 &$( - 7\, D^2 + 12\, D\, F - 9\, F^2) / 27$
                                 &$(11\, D^2 - 6\, D\, F - 9\, F^2) / 18$ \\     
              &$\Lambda\eta_{s}$   
                                 &$0$
                                 &$- (D + 3\, F)^2 / 27$
                                 &$(D + 3\, F)^2 / 27$
                                 &$(D + 3\, F)^2 / 27$
                                 &$0$ \\     
\noalign{\smallskip}
\hline
\noalign{\smallskip}
$\Xi^0$          &$\Xi \pi$      &$- (D - F)^2$
                                 &$(D - F)^2 / 3$
                                 &$- 4\, (D - F)^2 / 3$
                                 &$- 4\, (D - F)^2 / 3$
                                 &$0$ \\     
                 &$\Lambda K$    &$0$
                                 &$(D - 3\, F)^2 / 18$
                                 &$- (D - 3\, F)^2 / 18$
                                 &$(D - 3\, F)^2 / 9$
                                 &$- (D - 3\, F)^2 / 6$ \\     
                 &$\Sigma K$     &$(D + F)^2$
                                 &$(D + F)^2 / 6$
                                 &$5\, (D + F)^2 / 6$
                                 &$(D + F)^2 / 3$
                                 &$(D + F)^2 / 2$ \\     
                 &$\Omega K$     &$0$
                                 &$- (D - F)^2 / 3$
                                 &$(D - F)^2 / 3$
                                 &$- 2\, (D - F)^2 / 3$
                                 &$(D - F)^2$ \\     
                 &$\Xi \eta_{s}$     
                                 &$0$
                                 &$- 2\, (D^2 + 3\, F^2) / 9$
                                 &$2\, (D^2 + 3\, F^2) / 9$
                                 &$2\, (D^2 + 3\, F^2) / 9$
                                 &$0$ \\     
\noalign{\smallskip}	       
$\Xi^-$          &$\Xi \pi$      &$(D - F)^2$
                                 &$(D - F)^2 / 3$
                                 &$2\, (D - F)^2 / 3$
                                 &$2\, (D - F)^2 / 3$
                                 &$0$ \\     
                 &$\Lambda K$    &$(D - 3\, F)^2 / 6$
                                 &$(D - 3\, F)^2 / 18$
                                 &$(D - 3\, F)^2 / 9$
                                 &$(D - 3\, F)^2 / 9$
                                 &$0$ \\     
                 &$\Sigma K$     &$(D + F)^2 / 2$
                                 &$(D + F)^2 / 6$
                                 &$(D + F)^2 / 3$
                                 &$(D + F)^2 / 3$
                                 &$0$ \\     
                 &$\Omega K$     &$0$
                                 &$- (D - F)^2 / 3$
                                 &$(D - F)^2 / 3$
                                 &$(D - F)^2 / 3$
                                 &$0$ \\     
                 &$\Xi \eta_{s}$     
                                 &$0$
                                 &$- 2\, (D^2 + 3\, F^2) / 9$
                                 &$2\, (D^2 + 3\, F^2) / 9$
                                 &$2\, (D^2 + 3\, F^2) / 9$
                                 &$0$ \\     
\end{tabular}
\end{ruledtabular}
\end{table*}
%\hline
%\hline
%\end{longtable*}
\end{turnpage}

\begin{table*}[p]
\caption{Coefficients, $\chi$, providing the LNA contribution to
baryon magnetic moments by quark sectors with quark charges normalized
to unit charge.  Intermediate (Int.) meson-baryon channels are
indicated to allow for $SU(3)$-flavor breaking in both the meson and
baryon masses.  Total, direct sea-quark loop (Direct Loop), Valence,
indirect sea-quark loop (Indirect Loop) and Quenched Valence
coefficients are indicated.  The axial couplings take the tree-level
values $F=0.50$ and $D=0.76$ with $f_\pi = 93$ MeV.  Note $\epsilon =
0.0004$. }
\label{tab:quarkMomChi}
\addtolength{\tabcolsep}{-1pt}
\begin{ruledtabular}
\begin{tabular}{lcccccc}
$q$   &Int.           &Total     &Direct Loop &Valence  &Indirect Loop &Quenched Valence \\
\hline 
\noalign{\smallskip}
$u_p$ &$N \pi$        &$- 6.87$  &$+ 4.12$ &$- 11.0$ &$- 7.65$ &$- 3.33$ \\     
      &$\Lambda K$    &$- 3.68$  &$     0$ &$- 3.68$ &$- 3.68$ &$     0$ \\     
      &$\Sigma  K$    &$- 0.15$  &$     0$ &$- 0.15$ &$- 0.15$ &$     0$ \\     
\noalign{\smallskip}		           	               
$d_p$ &$N \pi$        &$+ 6.87$  &$+ 4.12$ &$+ 2.75$ &$- 0.59$ &$+ 3.33$ \\     
      &$\Sigma  K$    &$- 0.29$  &$     0$ &$- 0.29$ &$- 0.29$ &$     0$ \\     
\noalign{\smallskip}		           	               
$s_p$ &$\Lambda K$    &$+ 3.68$  &$+ 3.68$ &$     0$ &$     0$ &$     0$ \\     
      &$\Sigma K$     &$+ 0.44$  &$+ 0.44$ &$     0$ &$     0$ &$     0$ \\     
\noalign{\smallskip}
\hline 
\noalign{\smallskip}
$u_{\Sigma^+}$   &$\Sigma  \pi$  &$- 2.16$  &$+ 2.16$ &$- 4.32$ &$- 4.32$ &$     0$ \\     
                 &$\Lambda \pi$  &$- 1.67$  &$+ 1.67$ &$- 3.33$ &$- 3.33$ &$     0$ \\     
                 &$N K        $  &$     0$  &$+ 0.29$ &$- 0.29$ &$     0$ &$- 0.29$ \\     
                 &$\Xi K      $  &$- 6.87$  &$     0$ &$- 6.87$ &$- 3.83$ &$- 3.04$ \\     
\noalign{\smallskip}						                    
$d_{\Sigma^+}$   &$\Sigma  \pi$  &$+ 2.16$  &$+ 2.16$ &$     0$ &$     0$ &$     0$ \\     
                 &$\Lambda \pi$  &$+ 1.67$  &$+ 1.67$ &$     0$ &$     0$ &$     0$ \\     
                 &$N K        $  &$+ 0.29$  &$+ 0.29$ &$     0$ &$     0$ &$     0$ \\     
\noalign{\smallskip}
$s_{\Sigma^+}$   &$N K        $  &$- 0.29$  &$     0$ &$- 0.29$ &$- 0.59$ &$+ 0.29$ \\     
                 &$\Xi K      $  &$+ 6.87$  &$+ 3.83$ &$+ 3.04$ &$     0$ &$+ 3.04$ \\     
                 &$\Sigma \eta_{s}$  
                                 &$     0$  &$+ 0.29$ &$- 0.29$ &$- 0.29$ &$     0$ \\     
\noalign{\smallskip}
$u_{\Sigma^0} \mid d_{\Sigma^0}$   
                 &$\Sigma  \pi$  &$     0$  &$+ 2.16$ &$- 2.16$ &$- 2.16$ &$     0$ \\     
                 &$\Lambda \pi$  &$     0$  &$+ 1.67$ &$- 1.67$ &$- 1.67$ &$     0$ \\     
                 &$N K        $  &$+ 0.15$  &$+ 0.29$ &$- 0.15$ &$     0$ &$- 0.15$ \\     
                 &$\Xi K      $  &$- 3.43$  &$     0$ &$- 3.43$ &$- 1.91$ &$- 1.52$ \\     
\noalign{\smallskip}
\hline 
\noalign{\smallskip}
$u_\Lambda \mid d_\Lambda$
              &$\Sigma \pi $  &$     0$  &$+ 1.67$   &$- 1.67$    &$- 1.67$ &$     0$ \\    
              &$\Lambda\eta_{l}$
                              &$     0$  &$\epsilon$ &$-\epsilon$ &$-\epsilon$ &$     0$ \\
              &$N K        $  &$+ 3.68$  &$+ 2.45$   &$+ 1.23$    &$     0$ &$+ 1.23$ \\    
              &$\Xi K      $  &$- 0.40$  &$     0$   &$- 0.40$    &$- 0.83$ &$+ 0.44$ \\    
\noalign{\smallskip}						                      
$s_\Lambda$   &$\Lambda\eta_{s}$
                              &$     0$  &$+ 2.45$   &$- 2.45$    &$- 2.45$ &$     0$ \\    
              &$N K        $  &$- 7.36$  &$     0$   &$- 7.36$    &$- 4.91$ &$- 2.45$ \\    
              &$\Xi K      $  &$+ 0.79$  &$+ 1.67$   &$- 0.88$    &$     0$ &$- 0.88$ \\    
\noalign{\smallskip}
\hline
\noalign{\smallskip}
$u_{\Xi^0}$   &$\Xi \pi$      &$- 0.29$  &$+ 0.29$ &$- 0.59$ &$- 0.59$ &$     0$ \\     
              &$\Lambda K$    &$     0$  &$+ 0.40$ &$- 0.40$ &$     0$ &$- 0.40$ \\     
              &$\Sigma K$     &$+ 6.87$  &$+ 3.43$ &$+ 3.43$ &$     0$ &$+ 3.43$ \\     
              &$\Omega K$     &$     0$  &$     0$ &$     0$ &$- 0.29$ &$+ 0.29$ \\     
\noalign{\smallskip}					                         
$d_{\Xi^0}$   &$\Xi \pi$      &$+ 0.29$  &$+ 0.29$ &$     0$ &$     0$ &$     0$ \\     
              &$\Lambda K$    &$+ 0.40$  &$+ 0.40$ &$     0$ &$     0$ &$     0$ \\     
              &$\Sigma K$     &$+ 3.43$  &$+ 3.43$ &$     0$ &$     0$ &$     0$ \\     
\noalign{\smallskip}					                         
$s_{\Xi^0}$   &$\Lambda K$    &$- 0.40$  &$     0$ &$- 0.40$ &$- 0.79$ &$+ 0.40$ \\     
              &$\Sigma K$     &$- 10.3$  &$     0$ &$- 10.3$ &$- 6.87$ &$- 3.43$ \\     
              &$\Omega K$     &$     0$  &$+ 0.29$ &$- 0.29$ &$     0$ &$- 0.29$ \\     
              &$\Xi \eta_{s}$     
                              &$     0$  &$+ 3.83$ &$- 3.83$ &$- 3.83$ &$     0$ \\     
\end{tabular}
\end{ruledtabular}
\end{table*}

\begin{table*}[p]
%\squeezetable
\caption{Coefficients, $\chi$, providing the LNA contribution to
baryon magnetic moments.  Intermediate (Int.) meson-baryon channels
are indicated to allow for $SU(3)$-flavor breaking in both the meson
and baryon masses.  Total, direct sea-quark loop (Direct Loop),
Valence, indirect sea-quark loop (Indirect Loop) and Quenched Valence
coefficients are indicated.  The axial couplings take the tree-level
values $F=0.50$ and $D=0.76$ with $f_\pi = 93$ MeV.  Note $\epsilon =
0.0001$.}
\label{tab:baryonMomChi}
%\addtolength{\tabcolsep}{-1pt}
\begin{ruledtabular}
\begin{tabular}{ccccccc}
Baryon     &Channel        &Total     &Direct Loop &Valence  &Indirect
Loop  &Quenched Valence \\
\hline 
$p$        &$N \pi$        &$- 6.87$  &$+ 1.37$ &$- 8.24$ &$- 4.91$ &$- 3.33$ \\     
           &$\Lambda K$    &$- 3.68$  &$- 1.23$ &$- 2.45$ &$- 2.45$ &$     0$ \\     
           &$\Sigma  K$    &$- 0.15$  &$- 0.15$ &$     0$ &$     0$ &$     0$ \\     
\noalign{\bigskip}					                      
$n$        &$N \pi$        &$+ 6.87$  &$+ 1.37$ &$+ 5.49$ &$+ 2.16$ &$+ 3.33$ \\     
           &$\Lambda K$    &$     0$  &$- 1.23$ &$+ 1.23$ &$+ 1.23$ &$     0$ \\     
           &$\Sigma  K$    &$- 0.29$  &$- 0.15$ &$- 0.15$ &$- 0.15$ &$     0$ \\     
\noalign{\bigskip}					                      
$\Sigma^+$ &$\Sigma  \pi$  &$- 2.16$  &$+ 0.72$ &$- 2.88$ &$- 2.88$ &$     0$ \\     
           &$\Lambda \pi$  &$- 1.67$  &$+ 0.56$ &$- 2.22$ &$- 2.22$ &$     0$ \\     
           &$N K        $  &$     0$  &$+ 0.10$ &$- 0.10$ &$+ 0.20$ &$- 0.29$ \\     
           &$\Xi K      $  &$- 6.87$  &$- 1.28$ &$- 5.59$ &$- 2.55$ &$- 3.04$ \\     
           &$\Sigma \eta_{s}$
                           &$     0$  &$- 0.10$ &$+ 0.10$ &$+ 0.10$ &$     0$ \\     
\noalign{\bigskip}					                      
$\Sigma^0$ &$\Sigma  \pi$  &$     0$  &$+ 0.72$ &$- 0.72$ &$- 0.72$ &$     0$ \\     
           &$\Lambda \pi$  &$     0$  &$+ 0.56$ &$- 0.56$ &$- 0.56$ &$     0$ \\     
           &$N K        $  &$+ 0.15$  &$+ 0.10$ &$+ 0.05$ &$+ 0.20$ &$- 0.15$ \\     
           &$\Xi K      $  &$- 3.43$  &$- 1.28$ &$- 2.16$ &$- 0.64$ &$- 1.52$ \\     
           &$\Sigma \eta_{s}$
                           &$     0$  &$- 0.10$ &$+ 0.10$ &$+ 0.10$ &$     0$ \\     
\noalign{\bigskip}					                      
$\Sigma^-$ &$\Sigma  \pi$  &$+ 2.16$  &$+ 0.72$ &$+ 1.44$ &$+ 1.44$ &$     0$ \\     
           &$\Lambda \pi$  &$+ 1.67$  &$+ 0.56$ &$+ 1.11$ &$+ 1.11$ &$     0$ \\     
           &$N K        $  &$+ 0.29$  &$+ 0.10$ &$+ 0.20$ &$+ 0.20$ &$     0$ \\     
           &$\Xi K      $  &$     0$  &$- 1.28$ &$+ 1.28$ &$+ 1.28$ &$     0$ \\     
           &$\Sigma \eta_{s}$
                           &$     0$  &$- 0.10$ &$+ 0.10$ &$+ 0.10$ &$     0$ \\     
\noalign{\bigskip}
$\Lambda$  &$\Sigma \pi $  &$     0$  &$+ 0.56$   &$- 0.56$    &$- 0.56$ &$     0$ \\    
           &$\Lambda\eta_{l}$
                           &$     0$  &$\epsilon$ &$-\epsilon$ &$-\epsilon$ &$     0$ \\
           &$N K        $  &$+ 3.68$  &$+ 0.82$   &$+ 2.86$    &$+ 1.64$ &$+ 1.23$ \\    
           &$\Xi K      $  &$- 0.40$  &$- 0.56$   &$+ 0.16$    &$- 0.28$ &$+ 0.44$ \\    
           &$\Lambda\eta_{s}$
                           &$     0$  &$- 0.82$   &$+ 0.82$    &$+ 0.82$ &$     0$ \\    
\noalign{\smallskip}
$\Xi^0$    &$\Xi \pi$      &$- 0.29$  &$+ 0.10$ &$- 0.39$ &$- 0.39$ &$     0$ \\     
           &$\Lambda K$    &$     0$  &$+ 0.13$ &$- 0.13$ &$+ 0.26$ &$- 0.40$ \\     
           &$\Sigma K$     &$+ 6.87$  &$+ 1.14$ &$+ 5.72$ &$+ 2.29$ &$+ 3.43$ \\     
           &$\Omega K$     &$     0$  &$- 0.10$ &$+ 0.10$ &$- 0.20$ &$+ 0.29$ \\     
           &$\Xi    \eta_{s}$
                           &$     0$  &$- 1.28$ &$+ 1.28$ &$+ 1.28$ &$     0$ \\    
\noalign{\bigskip}					                      
$\Xi^-$    &$\Xi \pi$      &$+ 0.29$  &$+ 0.10$ &$+ 0.20$ &$+ 0.20$ &$     0$ \\     
           &$\Lambda K$    &$+ 0.40$  &$+ 0.13$ &$+ 0.26$ &$+ 0.26$ &$     0$ \\     
           &$\Sigma K$     &$+ 3.43$  &$+ 1.14$ &$+ 2.29$ &$+ 2.29$ &$     0$ \\     
           &$\Omega K$     &$     0$  &$- 0.10$ &$+ 0.10$ &$+ 0.10$ &$     0$ \\     
           &$\Xi    \eta_{s}$
                           &$     0$  &$- 1.28$ &$+ 1.28$ &$+ 1.28$ &$     0$ \\    
\end{tabular}
\end{ruledtabular}
\end{table*}

\end{section}

\begin{section}{Partial Quenching}
\label{sec:partQuench}

\begin{subsection}{Hadron Masses}

The flavor-blind nature of QCD makes it trivial to extend this
calculation of quenched baryon magnetic moments to the
partially-quenched theory.  As new flavors are introduced through the
use of dynamically generated gauge fields, one simply adds the direct
and indirect sea-quark loop contributions evaluated in
Sec.~\ref{sec:mom} to the quenched results of
Sec.~\ref{sec:results}.  To incorporate hadron mass violations of
$SU(3)$-flavor symmetry, one must track the meson mass of the
valence-sea meson.  As we have already isolated each valence quark
flavor contribution to the baryon moment, the mass of the meson is
identified by the valence- and sea-quark mass composing the meson.

It should be noted that the double hair-pin graph of the $\eta'$ meson
remains anomalous in the partially-quenched theory \cite{Sharpe:2001fh}.
However, the contribution of the $\eta'$ propagator is suppressed by
the difference in valence- and sea-quark masses.

\end{subsection}

\begin{subsection}{Sea- and Ghost-Quark Electric Charge Assignments}

There has been some discussion on the electric charge assignments that
may be applied to the various quark sectors of partially-quenched
effective field theory \cite{Chen:2001yi}.  In the conventional view
of quenched chiral perturbation theory, the charges of the commuting
ghost-quark fields are tied to the valence quark charges in order to
eliminate both the direct and indirect sea-quark loop contributions of
the valence sector.  Similarly, for partially-quenched chiral
perturbation theory, it is usually argued that the ghost quarks are
identical to the valence quarks, except for their statistics.

However, it has been indicated that when the number of sea quarks
matches the number of valence quarks, more general charge assignments
are possible \cite{Chen:2001yi}.  The idea is that when the masses and
charges of the sea- and ghost-quarks match, these contributions cancel
leaving the theory of full QCD.  In this case the charges of the sea
and ghost quarks need not be related to the the valence quarks.
However, it is essential that the quark masses of the valence and ghost
sectors match, such that the indirect sea-quark loop contributions of
the valence sector continue to be quenched.

We have already argued in the Introduction that it is important to
provide an opportunity to include disconnected insertions of the
electromagnetic current in the quenched approximation.  These
insertions can be calculated in the quenched approximation and give
rise to direct sea-quark loop contributions.  It is now clear that
this goal can be realized in the formal theory of quenched chiral
perturbation theory by assigning neutral electric charges to the
ghost-quark fields.  Indirect sea-quark loop contributions are removed
while leaving direct sea-quark loop contributions from the valence
sector unaltered.

\end{subsection}

\begin{subsection}{Examples}

Consider for example the quark sector contributions to a baryon
magnetic moment in a partially quenched theory with two degenerate
light quarks and one heavy sea quark, labeled $u'$, $d'$ and $s'$.
Electric charge assignments are $q_u$, $q_d$, $q_s$ for the valence
sector of the theory and $q_u^\prime$, $q_d^\prime$, $q_s^\prime$ for
the ghost- and sea-quark sectors.

\begin{subsubsection}{Proton Magnetic Moment}

The quenched quark-sector results for the proton are complemented by
direct sea-quark loop contributions from the valence- and ghost-quark
sectors plus both direct and indirect contributions from the sea-quark
sector.  As discussed in Sec.~\ref{subsec:proton} such loop
contributions are flavor blind and the couplings are easily extracted
from Tables \ref{tab:uProtonSea}, \ref{tab:dProtonSea} or
\ref{tab:sProtonTotal} for the direct contributions and Table
\ref{tab:uProtonQuench} for the indirect contribution.  For
simplicity, we will suppress baryon mass splittings in the following.
However, they may be introduced in a transparent manner.

For the $u$-quark sector in the proton, one has
\begin{eqnarray}
u_p &=& \xi \left \{ \,
- q_u\, {4 \over 3} D^2 \, m_\pi 
+ q_u\, {1 \over 3} (5\, D^2 - 6\, D\, F + 9\, F^2) \, m_\pi  
\right . \nonumber \\
&& \qquad 
+ q_u^\prime\, {1 \over 3} (5\, D^2 - 6\, D\, F + 9\, F^2) \, 
\left ( \widetilde m_\pi - m_\pi \right )
\nonumber \\
&& \qquad 
- q_u\, {4 \over 3} (D^2 + 3\, F^2)\, \widetilde m_\pi
\nonumber \\
&& \qquad \left .
- q_u\, {2 \over 3} (D^2 + 3\, F^2)\, \widetilde m_K 
\right \} \, , 
\end{eqnarray}
where $\xi \equiv {m_N / ( 8 \pi f_\pi^2) }$.  $\widetilde m_\pi$
denotes a $\pi$-meson composed of a light-valence and light-sea quark,
and $\widetilde m_K$ denotes a $K$-meson composed of a light-valence
and heavy-sea quark.  The second term is a direct $u$ sea-quark loop
contribution associated with the valence sector, canceled by the
ghost-quark contribution in the third term when $q_u^\prime = q_u$.  The
third term also includes the direct $u'$ sea-quark loop contribution
associated with the sea-quark sector and originates from
Table~\ref{tab:uProtonSea} for Figs.~\ref{ProtonCloud}(b) and (e).
The last two terms are indirect $u$ sea-quark loop contributions and
originate from Table~\ref{tab:uProtonQuench} for diagrams (b), (g) and
(h) respectively.

Similarly
\begin{eqnarray}
d_p &=& \xi \left \{ \,
+ q_d\, {4 \over 3} D^2 \, m_\pi 
+ q_d\, {1 \over 3} (5\, D^2 - 6\, D\, F + 9\, F^2) \, m_\pi  
\right . \nonumber \\
&& \qquad 
+ q_d^\prime\, {1 \over 3} (5\, D^2 - 6\, D\, F + 9\, F^2) \, 
\left ( \widetilde m_\pi - m_\pi \right ) 
\nonumber \\
&& \qquad 
- q_d\, 2\, (D - F)^2 \, \widetilde m_\pi
- q_d\, (D - F)^2 \, \widetilde m_K 
\biggr \} \, , 
\nonumber \\
\end{eqnarray}
and
\begin{eqnarray}
s_p &=& \xi \left \{ \,
q_s\, {1 \over 3} (5\, D^2 - 6\, D\, F + 9\, F^2) \, m_K
\right . \\
&& \qquad \left .
+ q_s^\prime\, {1 \over 3} (5\, D^2 - 6\, D\, F + 9\, F^2) \, 
\left ( \widetilde m_K - m_K \right )
\right \} \, . 
\nonumber
\end{eqnarray}
The LNA behavior of the proton magnetic moment in the partially
quenched theory is
\begin{eqnarray}
\mu_p &=& \xi \left \{ \,
- {4 \over 3} D^2 \, m_\pi 
\right . \nonumber \\
&& \quad 
+ {1 \over 9} (5\, D^2 - 6\, D\, F + 9\, F^2) \, 
\left ( m_\pi - m_K \right )
\nonumber \\
&& \quad 
+ \left ( q_u^\prime + q_d^\prime \right ) \, 
{1 \over 3} (5\, D^2 - 6\, D\, F + 9\, F^2) \, 
\left ( \widetilde m_\pi - m_\pi \right )
\nonumber \\
&& \quad 
+ q_s^\prime\, {1 \over 3} (5\, D^2 - 6\, D\, F + 9\, F^2) \, 
\left ( \widetilde m_K - m_K \right )
\nonumber \\
&& \quad \left .
- {2 \over 9} (D + 3\, F)^2 \, \widetilde m_\pi
- {1 \over 9} (D + 3\, F)^2 \, \widetilde m_K
\right \} \, . 
\end{eqnarray}
We note that this final expression agrees with that of Eq.~(48) in
Ref.~\cite{Chen:2001yi}.

\end{subsubsection}

\begin{subsubsection}{$\Sigma^+$ Magnetic Moment}

To clearly establish the method for constructing partially-quenched
chiral coefficients, we consider the $\Sigma^+$ hyperon.  The direct
and indirect sea-quark loop contributions are flavor blind and the
couplings are easily extracted from Tables \ref{tab:uSigmaSea},
\ref{tab:dSigmaTotal} or \ref{tab:sSigmaSea} for the direct
contributions and Table \ref{tab:uSigmaQuench} for the indirect
contribution.  For the $u$-quark sector in $\Sigma^+$, one has
\begin{eqnarray}
u_{\Sigma^+} &=& \xi \left \{ \,
- q_u\, {4 \over 3} D^2 \, m_K 
\right . 
\nonumber \\
&& \qquad 
+ q_u\, {2 \over 3} (D^2 + 3\, F^2) \, m_\pi  
+ q_u\, (D - F)^2 \, m_K  
\nonumber \\
&& \qquad 
+ q_u^\prime\, {2 \over 3} (D^2 + 3\, F^2) \, 
\left ( \widetilde m_\pi - m_\pi \right )
\nonumber \\
&& \qquad 
+ q_u^\prime\, (D - F)^2 \, 
\left ( \widehat m_K - m_K \right )
\nonumber \\
&& \qquad 
- q_u\, {4 \over 3} (D^2 + 3\, F^2) \, \widetilde m_\pi  
\nonumber \\
&& \qquad \left .
- q_u\, {2 \over 3} (D^2 + 3\, F^2) \, \widetilde m_K  
\right \} \, , 
\end{eqnarray}
where $\widetilde m_K$ denotes a $K$-meson composed of a light-valence
and heavy-sea quark and $\widehat m_K$ denotes a $K$-meson composed of
a strange-valence and light-sea quark.  The second and third terms are
direct $u$ sea-quark loop contributions associated with the valence
sector, canceled by the ghost-quark contribution in the fourth and
fifth terms when $q_u^\prime = q_u$.  The fourth and fifth terms also
include the direct $u'$ sea-quark loop contribution associated with
the sea-quark sector and originate from Table~\ref{tab:uSigmaSea} for
Figs.~\ref{SigmaCloud}(b) and (e).  The last two terms are indirect
$u$ sea-quark loop contributions and originate from
Table~\ref{tab:uSigmaQuench} for Figs.~\ref{SigmaCloud}$(b)+(c)$ and
(h) respectively.  Similarly
\begin{eqnarray}
s_{\Sigma^+} &=& \xi \left \{ \,
+ q_s\, {4 \over 3} D^2 \, m_K 
\right . \\
&& \qquad
+ q_s\, {2 \over 3} (D^2 + 3\, F^2) \, m_K  
+ q_s\, (D - F)^2 \, m_{\eta_s}
\nonumber \\
&& \qquad 
+ q_s^\prime\, {2 \over 3} (D^2 + 3\, F^2) \, 
\left ( \widetilde m_K -m_K \right )
\nonumber \\
&& \qquad 
+ q_s^\prime\, (D - F)^2 \, 
\left ( \widetilde m_{\eta_s} - m_{\eta_s} \right )
\nonumber \\
&& \qquad %\left .
- q_s\, 2\, (D - F)^2 \, \widehat m_K
- q_s\, (D - F)^2 \, \widetilde m_{\eta_s}
\biggr \} \, , 
\nonumber
\end{eqnarray}
where $\widetilde m_{\eta_s}$ denotes an $s\, \overline s^{\,\prime}$
$\eta$-meson composed of a strange-valence and anti-heavy sea-quark,
and
\begin{eqnarray}
d_{\Sigma^+} &=& \xi \left \{ \,
+ q_d\, {2 \over 3} (D^2 + 3\, F^2) \, m_\pi  
+ q_d\, (D - F)^2 \, m_K  
\right . \nonumber \\
&& \qquad
+ q_d^\prime\, {2 \over 3} (D^2 + 3\, F^2) \, 
\left ( \widetilde m_\pi - m_\pi \right )
\nonumber \\
&& \qquad
+ q_d^\prime\, (D - F)^2 \, 
\left ( \widehat m_K - m_K \right )
\biggr \} \, .
\end{eqnarray}
Thus, the LNA behavior of the $\Sigma^+$ magnetic moment in the
partially quenched theory is
\begin{eqnarray}
\mu_{\Sigma^+} &=& \xi \left \{ \,
- {4 \over 3} D^2 \, m_K 
\right . \nonumber \\
&& \qquad 
+ {2 \over 9} (D^2 + 3\, F^2) \, m_\pi  
+ {1 \over 3} (D - F)^2 \, m_K  
\nonumber \\
&& \qquad
- {2 \over 9} (D^2 + 3\, F^2) \, m_K  
- {1 \over 3} (D - F)^2 \, m_{\eta_s}
\nonumber \\
&& \qquad 
+ \left ( q_u^\prime + q_d^\prime \right ) 
{2 \over 3} (D^2 + 3\, F^2) \, 
\left ( \widetilde m_\pi - m_\pi \right )
\nonumber \\
&& \qquad 
+ \left ( q_u^\prime + q_d^\prime \right )
 (D - F)^2 \, 
\left ( \widehat m_K - m_K \right )
\nonumber \\
&& \qquad 
+ q_s^\prime\, {2 \over 3} (D^2 + 3\, F^2) \, 
\left ( \widetilde m_K -m_K \right )
\nonumber \\
&& \qquad 
+ q_s^\prime\, (D - F)^2 \, 
\left ( \widetilde m_{\eta_s} - m_{\eta_s} \right )
\nonumber \\
&& \qquad 
+ {1 \over 3} (D - F)^2 \, 
\left ( 2\, \widehat m_K + \widetilde m_{\eta_s} \right )
\nonumber \\
&& \qquad \left .
- {4 \over 9} (D^2 + 3\, F^2) \, 
\left ( 2\, \widetilde m_\pi + \widetilde m_K \right )
\right \} \, ,
\end{eqnarray}
again in agreement with Eq.~(51) of Ref.~\cite{Chen:2001yi}.

\end{subsubsection}

Partially-quenched results may be similarly obtained for the remainder
of the quark-sector contributions to octet baryon magnetic moments
using the approach described here in detail.  Since the results
require specific knowledge of the number and nature of dynamical
flavors, we defer writing further specific results.

\end{subsection}
\end{section}

\begin{table*}[p]
\caption{One-loop corrected coefficients, $\chi$, providing the LNA
contribution to baryon magnetic moments by quark sectors with quark
charges normalized to unit charge.  Total, direct sea-quark loop
(Direct Loop), Valence, indirect sea-quark loop (Indirect Loop) and
Quenched Valence coefficients are indicated.  Here, the axial
couplings take the one-loop corrected values $F=0.40$ and $D=0.61$
with $f_\pi = 93$ MeV.  Note $\epsilon = 0.0004$. }
\label{tab:quarkMomChiOneLoop}
\addtolength{\tabcolsep}{-1pt}
\begin{ruledtabular}
\begin{tabular}{lcccccc}
$q$   &Int.           &Total     &Direct Loop &Valence  &Indirect Loop &Quenched Valence \\
\hline 
\noalign{\smallskip}
$u_p$ &$N \pi$        &$- 4.41$  &$+ 2.65$ &$- 7.06$ &$- 4.91$ &$- 2.15$ \\     
      &$\Lambda K$    &$- 2.36$  &$     0$ &$- 2.36$ &$- 2.36$ &$     0$ \\     
      &$\Sigma  K$    &$- 0.10$  &$     0$ &$- 0.10$ &$- 0.10$ &$     0$ \\     
\noalign{\smallskip}		           	               
$d_p$ &$N \pi$        &$+ 4.41$  &$+ 2.65$ &$+ 1.76$ &$- 0.38$ &$+ 2.15$ \\     
      &$\Sigma  K$    &$- 0.19$  &$     0$ &$- 0.19$ &$- 0.19$ &$     0$ \\     
\noalign{\smallskip}		           	               
$s_p$ &$\Lambda K$    &$+ 2.36$  &$+ 2.36$ &$     0$ &$     0$ &$     0$ \\     
      &$\Sigma K$     &$+ 0.29$  &$+ 0.29$ &$     0$ &$     0$ &$     0$ \\     
\noalign{\smallskip}
\hline 
\noalign{\smallskip}
$u_{\Sigma^+}$   &$\Sigma  \pi$  &$- 1.38$  &$+ 1.38$ &$- 2.77$ &$- 2.77$ &$     0$ \\     
                 &$\Lambda \pi$  &$- 1.07$  &$+ 1.07$ &$- 2.15$ &$- 2.15$ &$     0$ \\     
                 &$N K        $  &$     0$  &$+ 0.19$ &$- 0.19$ &$     0$ &$- 0.19$ \\     
                 &$\Xi K      $  &$- 4.41$  &$     0$ &$- 4.41$ &$- 2.46$ &$- 1.95$ \\     
\noalign{\smallskip}						                    
$d_{\Sigma^+}$   &$\Sigma  \pi$  &$+ 1.38$  &$+ 1.38$ &$     0$ &$     0$ &$     0$ \\     
                 &$\Lambda \pi$  &$+ 1.07$  &$+ 1.07$ &$     0$ &$     0$ &$     0$ \\     
                 &$N K        $  &$+ 0.19$  &$+ 0.19$ &$     0$ &$     0$ &$     0$ \\     
\noalign{\smallskip}
$s_{\Sigma^+}$   &$N K        $  &$- 0.19$  &$     0$ &$- 0.19$ &$- 0.38$ &$+ 0.19$ \\     
                 &$\Xi K      $  &$+ 4.41$  &$+ 2.46$ &$+ 1.95$ &$     0$ &$+ 1.95$ \\     
                 &$\Sigma \eta_{s}$  
                                 &$     0$  &$+ 0.19$ &$- 0.19$ &$- 0.19$ &$     0$ \\     
\noalign{\smallskip}
$u_{\Sigma^0} \mid d_{\Sigma^0}$   
                 &$\Sigma  \pi$  &$     0$  &$+ 1.38$ &$- 1.38$ &$- 1.38$ &$     0$ \\     
                 &$\Lambda \pi$  &$     0$  &$+ 1.07$ &$- 1.07$ &$- 1.07$ &$     0$ \\     
                 &$N K        $  &$+ 0.10$  &$+ 0.19$ &$- 0.10$ &$     0$ &$- 0.10$ \\     
                 &$\Xi K      $  &$- 2.21$  &$     0$ &$- 2.21$ &$- 1.23$ &$- 0.98$ \\     
\noalign{\smallskip}
\hline 
\noalign{\smallskip}
$u_\Lambda \mid d_\Lambda$
              &$\Sigma \pi $  &$     0$  &$+ 1.07$   &$- 1.07$    &$- 1.07$ &$     0$ \\    
              &$\Lambda\eta_{l}$
                              &$     0$  &$\epsilon$ &$-\epsilon$ &$-\epsilon$ &$     0$ \\
              &$N K        $  &$+ 2.36$  &$+ 1.57$   &$+ 0.79$    &$     0$ &$+ 0.79$ \\    
              &$\Xi K      $  &$- 0.25$  &$     0$   &$- 0.25$    &$- 0.54$ &$+ 0.29$ \\    
\noalign{\smallskip}						                      
$s_\Lambda$   &$\Lambda\eta_{s}$
                              &$     0$  &$+ 1.57$   &$- 1.57$    &$- 1.57$ &$     0$ \\    
              &$N K        $  &$- 4.72$  &$     0$   &$- 4.72$    &$- 3.15$ &$- 1.57$ \\    
              &$\Xi K      $  &$+ 0.50$  &$+ 1.07$   &$- 0.57$    &$     0$ &$- 0.57$ \\    
\noalign{\smallskip}
\hline
\noalign{\smallskip}
$u_{\Xi^0}$   &$\Xi \pi$      &$- 0.19$  &$+ 0.19$ &$- 0.38$ &$- 0.38$ &$     0$ \\     
              &$\Lambda K$    &$     0$  &$+ 0.25$ &$- 0.25$ &$     0$ &$- 0.25$ \\     
              &$\Sigma K$     &$+ 4.41$  &$+ 2.21$ &$+ 2.21$ &$     0$ &$+ 2.21$ \\     
              &$\Omega K$     &$     0$  &$     0$ &$     0$ &$- 0.19$ &$+ 0.19$ \\     
\noalign{\smallskip}					                         
$d_{\Xi^0}$   &$\Xi \pi$      &$+ 0.19$  &$+ 0.19$ &$     0$ &$     0$ &$     0$ \\     
              &$\Lambda K$    &$+ 0.25$  &$+ 0.25$ &$     0$ &$     0$ &$     0$ \\     
              &$\Sigma K$     &$+ 2.21$  &$+ 2.21$ &$     0$ &$     0$ &$     0$ \\     
\noalign{\smallskip}					                         
$s_{\Xi^0}$   &$\Lambda K$    &$- 0.25$  &$     0$ &$- 0.25$ &$- 0.50$ &$+ 0.25$ \\     
              &$\Sigma K$     &$- 6.62$  &$     0$ &$- 6.62$ &$- 4.41$ &$- 2.21$ \\     
              &$\Omega K$     &$     0$  &$+ 0.19$ &$- 0.19$ &$     0$ &$- 0.19$ \\     
              &$\Xi \eta_{s}$     
                              &$     0$  &$+ 2.46$ &$- 2.46$ &$- 2.46$ &$     0$ \\     
\end{tabular}
\end{ruledtabular}
\end{table*}

\begin{table*}[p]
%\squeezetable
\caption{One-loop corrected coefficients, $\chi$, providing the LNA
contribution to baryon magnetic moments.  Total, direct sea-quark loop
(Direct Loop), Valence, indirect sea-quark loop (Indirect Loop) and
Quenched Valence coefficients are indicated.  Here, the axial
couplings take the one-loop corrected values $F=0.40$ and $D=0.61$
with $f_\pi = 93$ MeV.  Note $\epsilon = 0.000128$.}
\label{tab:baryonMomChiOneLoop}
%\addtolength{\tabcolsep}{-1pt}
\begin{ruledtabular}
\begin{tabular}{ccccccc}
Baryon     &Channel        &Total     &Direct Loop &Valence  &Indirect
Loop  &Quenched Valence \\
\hline 
$p$        &$N \pi$        &$- 4.41$  &$+ 0.88$ &$- 5.29$ &$- 3.15$ &$- 2.15$ \\     
           &$\Lambda K$    &$- 2.36$  &$- 0.79$ &$- 1.57$ &$- 1.57$ &$     0$ \\     
           &$\Sigma  K$    &$- 0.10$  &$- 0.10$ &$     0$ &$     0$ &$     0$ \\     
\noalign{\bigskip}					                      
$n$        &$N \pi$        &$+ 4.41$  &$+ 0.88$ &$+ 3.53$ &$+ 1.38$ &$+ 2.15$ \\     
           &$\Lambda K$    &$     0$  &$- 0.79$ &$+ 0.79$ &$+ 0.79$ &$     0$ \\     
           &$\Sigma  K$    &$- 0.19$  &$- 0.10$ &$- 0.10$ &$- 0.10$ &$     0$ \\     
\noalign{\bigskip}					                      
$\Sigma^+$ &$\Sigma  \pi$  &$- 1.38$  &$+ 0.46$ &$- 1.85$ &$- 1.85$ &$     0$ \\     
           &$\Lambda \pi$  &$- 1.07$  &$+ 0.36$ &$- 1.43$ &$- 1.43$ &$     0$ \\     
           &$N K        $  &$     0$  &$+ 0.06$ &$- 0.06$ &$+ 0.13$ &$- 0.19$ \\     
           &$\Xi K      $  &$- 4.41$  &$- 0.82$ &$- 3.59$ &$- 1.64$ &$- 1.95$ \\     
           &$\Sigma \eta_{s}$
                           &$     0$  &$- 0.06$ &$+ 0.06$ &$+ 0.06$ &$     0$ \\     
\noalign{\bigskip}					                      
$\Sigma^0$ &$\Sigma  \pi$  &$     0$  &$+ 0.46$ &$- 0.46$ &$- 0.46$ &$     0$ \\     
           &$\Lambda \pi$  &$     0$  &$+ 0.36$ &$- 0.36$ &$- 0.36$ &$     0$ \\     
           &$N K        $  &$+ 0.10$  &$+ 0.06$ &$+ 0.03$ &$+ 0.13$ &$- 0.10$ \\     
           &$\Xi K      $  &$- 2.21$  &$- 0.82$ &$- 1.39$ &$- 0.41$ &$- 0.98$ \\     
           &$\Sigma \eta_{s}$
                           &$     0$  &$- 0.06$ &$+ 0.06$ &$+ 0.06$ &$     0$ \\     
\noalign{\bigskip}					                      
$\Sigma^-$ &$\Sigma  \pi$  &$+ 1.38$  &$+ 0.46$ &$+ 0.92$ &$+ 0.92$ &$     0$ \\     
           &$\Lambda \pi$  &$+ 1.07$  &$+ 0.36$ &$+ 0.72$ &$+ 0.72$ &$     0$ \\     
           &$N K        $  &$+ 0.19$  &$+ 0.06$ &$+ 0.13$ &$+ 0.13$ &$     0$ \\     
           &$\Xi K      $  &$     0$  &$- 0.82$ &$+ 0.82$ &$+ 0.82$ &$     0$ \\     
           &$\Sigma \eta_{s}$
                           &$     0$  &$- 0.06$ &$+ 0.06$ &$+ 0.06$ &$     0$ \\     
\noalign{\bigskip}
$\Lambda$  &$\Sigma \pi $  &$     0$  &$+ 0.36$   &$- 0.36$    &$- 0.36$ &$     0$ \\    
           &$\Lambda\eta_{l}$
                           &$     0$  &$\epsilon$ &$-\epsilon$ &$-\epsilon$ &$     0$ \\
           &$N K        $  &$+ 2.36$  &$+ 0.53$   &$+ 1.84$    &$+ 1.05$ &$+ 0.79$ \\    
           &$\Xi K      $  &$- 0.25$  &$- 0.36$   &$+ 0.11$    &$- 0.18$ &$+ 0.29$ \\    
           &$\Lambda\eta_{s}$
                           &$     0$  &$- 0.53$   &$+ 0.53$    &$+ 0.53$ &$     0$ \\    
\noalign{\smallskip}
$\Xi^0$    &$\Xi \pi$      &$- 0.19$  &$+ 0.06$ &$- 0.25$ &$- 0.25$ &$     0$ \\     
           &$\Lambda K$    &$     0$  &$+ 0.08$ &$- 0.08$ &$+ 0.17$ &$- 0.25$ \\     
           &$\Sigma K$     &$+ 4.41$  &$+ 0.74$ &$+ 3.68$ &$+ 1.47$ &$+ 2.21$ \\     
           &$\Omega K$     &$     0$  &$- 0.06$ &$+ 0.06$ &$- 0.13$ &$+ 0.19$ \\     
           &$\Xi    \eta_{s}$
                           &$     0$  &$- 0.82$ &$+ 0.82$ &$+ 0.82$ &$     0$ \\    
\noalign{\bigskip}					                      
$\Xi^-$    &$\Xi \pi$      &$+ 0.19$  &$+ 0.06$ &$+ 0.13$ &$+ 0.13$ &$     0$ \\     
           &$\Lambda K$    &$+ 0.25$  &$+ 0.08$ &$+ 0.17$ &$+ 0.17$ &$     0$ \\     
           &$\Sigma K$     &$+ 2.21$  &$+ 0.74$ &$+ 1.47$ &$+ 1.47$ &$     0$ \\     
           &$\Omega K$     &$     0$  &$- 0.06$ &$+ 0.06$ &$+ 0.06$ &$     0$ \\     
           &$\Xi    \eta_{s}$
                           &$     0$  &$- 0.82$ &$+ 0.82$ &$+ 0.82$ &$     0$ \\    
\end{tabular}
\end{ruledtabular}
\end{table*}

\begin{section}{Summary}
\label{sec:summary}

The diagrammatic method for separating valence and sea-quark-loop
contributions to the meson cloud of hadrons provides a transparent
approach to the calculation of quenched chiral coefficients.  The
origin of chiral nonanalytic structure is obvious, and facilitates the
incorporation of the correct nonanalytic structure matching today's
numerical simulations.  In the process, the coefficients for
partially-quenched QCD are derived; no new calculations are required.

The valence sector of full QCD contains the largest coefficients for
the leading nonanalytic behavior of magnetic moments.  The $u$-quark
contribution to the proton magnetic moment has a coefficient of
$-11.0$ for the rapidly varying pion-cloud contribution, which is
complemented further by the kaon cloud.  These are connected
insertions of the electromagnetic current in full QCD and should
reveal significant curvature in the approach to the chiral limit.  It
is also encouraging to note that the $u$-quark sector is known to have
relatively small statistical uncertainties in the quenched
approximation \cite{Leinweber:1990dv} compared to that for the $d$
quark.

The coefficients of the leading nonanalytic terms of full QCD change
significantly upon quenching.  Some channels still hold excellent
promise for revealing the nonanalytic behavior of meson-cloud physics
even in the quenched approximation.  For example, the $u$ or $d$-quark
in the proton have large coefficients for the nonanalytic term
proportional to $m_\pi$, with opposite signs respectively.  Similarly,
both the proton and neutron magnetic moments have large coefficients
surviving in quenched QCD.  Because the $u$-quark in the proton has
significantly smaller statistical errors than that for the $d$ quark
in the proton \cite{Leinweber:1990dv}, the $u$-quark contribution to
the proton magnetic moment provides the optimal opportunity to directly
view nonanalytic behavior associated with the quenched meson cloud of
baryons in the quenched approximation.  Figure~\ref{uQuarkP}
illustrates the anticipated curvature \cite{Leinweber:1998ej}
associated with the term $-(4/3)\, D^2 \, m_N \, m_\pi / (8\, \pi \,
f_\pi^2)$ surviving in the quenched approximation.

\begin{figure}[t]
\centering{\
\epsfig{file=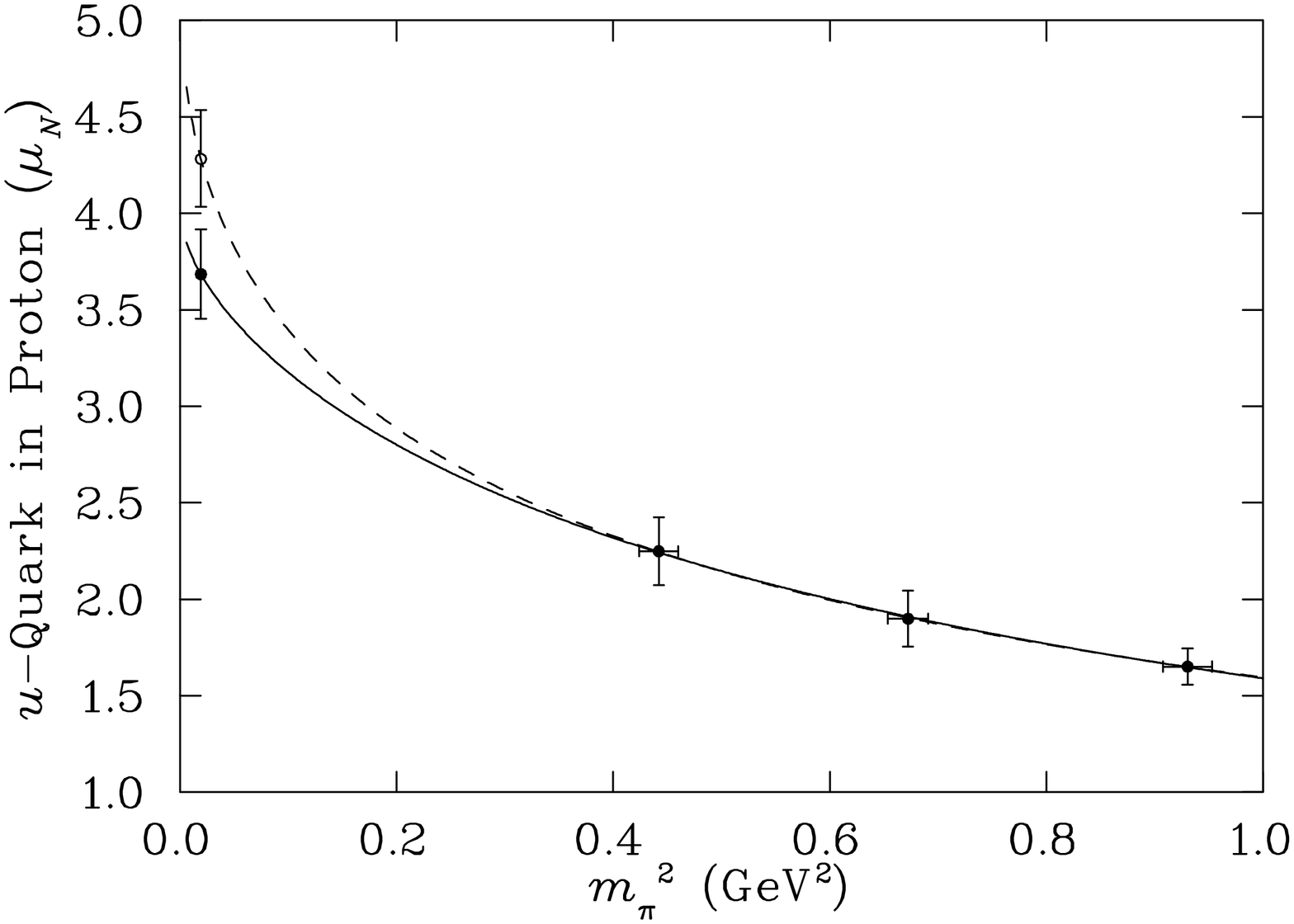,height=\hsize,angle=90} }
\caption{Chiral extrapolation via the Pad\'e of
\protect\cite{Leinweber:1998ej,Hackett-Jones:2000js}.  The solid curve
displays the self-consistent extrapolation of the quenched simulation
results of \protect\cite{Leinweber:1990dv}, whereas the dashed curve
shows the curvature of full QCD.  One-loop corrected axial couplings
are used in the Pad\'e.}
\label{uQuarkP}
\end{figure}

There are other interesting opportunities.  Consider for example the
$s$-quark contribution to the quenched $\Lambda$ magnetic moment.
Table~\ref{tab:quarkMomChi} indicates that the coefficient of the $NK$
contribution to the $\Lambda$ magnetic moment is large.  Because the
nucleon is significantly lighter than the $\Lambda$, the $N K$ loop
can contribute enhanced nonlinear behavior.  Hence, there is a
prediction of significant curvature in the extrapolation of the
$s$-quark contribution, even when the mass of the strange quark is
held fixed as is commonly done in lattice QCD simulations.  As such,
the effect is purely environmental
\cite{Leinweber:1990dv,Leinweber:1992hy,Leinweber:1995ie,Leinweber:1999nf}.
The effect arises from the extrapolation of light $u$ and $d$ quarks
in the $\Lambda$.

The $s$-quark in $\Xi$ provides another opportunity to observe a
purely environmental effect in quenched QCD.  Here the coefficient of
the $\Sigma K$ channel is very large, again predicting curvature in
the $s$-quark contribution to the magnetic moment, even when the
$s$-quark mass is held fixed.  The mass of $\Sigma$ is less than the
mass of $\Xi$ allowing the kaon to provide enhanced nonlinear
behavior.

A few channels hold potential for revealing {\it quenched artifacts}
in quenched simulations.  Despite the prediction of curvature for both
the $s$ and $d$-quark sectors of $\Xi^-$ in both full and quenched
QCD, the extrapolation of the total $\Xi^-$-baryon magnetic moment
receives no leading nonanalytic contribution from neither the $\pi$-
nor the $K$-meson cloud in the quenched approximation.  Similar
results hold for the $\Sigma^-$ magnetic moment.

It is particularly difficult to directly determine the loop
contribution to baryon magnetic moments in numerical simulations
\cite{Dong:1997xr,Mathur:2000cf,Wilcox:2000qa,Lewis:2002ix}.  As such,
it is of particular interest to compare the coefficients of the
valence quark contributions in full QCD (column ``Valence'' of Tables
\ref{tab:quarkMomChi} and \ref{tab:baryonMomChi}) to that for the
valence quark contributions of quenched QCD (column ``Quenched
Valence'' of Tables \ref{tab:quarkMomChi} and \ref{tab:baryonMomChi}).
Here, the $u$-quark in $\Sigma^+$ stands out with the significant
curvature of the $\Sigma \pi$ and $\Lambda \pi$ channels completely
suppressed from -4.32 to 0 and -3.33 to 0 respectively.  Only the $\Xi
K$ channel has a significant coupling for the $u$ quark in the
quenched $\Sigma^+$, but curvature in this channel is suppressed by
the large excitation energy required to form the intermediate state.
The $u$ quark in the proton is also worthy of note, with the
coefficient of the rapidly-varying $\pi N$ channel dropping
significantly from $-11.0$ in full QCD to $-3.33$ in quenched QCD and
the kaon contribution vanishing completely.

In summary, this study of quark-sector contributions to baryon
magnetic moments in quenched, partially-quenched and full QCD
indicates there are numerous opportunities to observe and
understand the underlying structure of baryons and the nature of
chiral nonanalytic behavior in QCD and its quenched variants.
Numerical simulations of the observables discussed herein are
currently in production on the Australian Partnership for Advanced
Computing (APAC) National Facility using FLIC fermions
\cite{Zanotti:2001yb} which provide efficient access to the light
quark-mass regime.  It will be interesting to confront these
predictions with numerical simulation results.

\end{section}

%%%%%%%%%%%%%%%%%%%%%%%%%%%%%%%%%%%%%%%%%%%%%%%%%%%%%%%%%%%%%%%%%%%%%%%%%%%

\acknowledgments

Thanks to Matthias Burkardt, Ian Cloet, Ben Crouch, Martin Savage,
Tony Thomas, Tony Williams, Stewart Wright and Ross Young for
beneficial discussions.  This research is supported by the Australian
Research Council.

%%%%%%%%%%%%%%%%%%%%%%%%%%%%%%%%%%%%%%%%%%%%%%%%%%%%%%%%%%%%%%%%%%%%%%%%%%%

\end{document}